\documentclass[aps,prl,twocolumn,superscriptaddress]{revtex4-2}
\usepackage{bm}
\usepackage{graphicx}
\usepackage{color}
\usepackage{braket}
\usepackage{amsmath,amssymb,amsfonts,amsthm,mathtools}
\usepackage{enumerate}
\usepackage{soul}
\usepackage{enumitem}
\usepackage{subfigure}
\usepackage[colorlinks=true,linkcolor=blue,anchorcolor=red,citecolor=blue,urlcolor=blue]{hyperref}
\usepackage{titlesec}
\usepackage{tikz-cd} 
\usepackage{float}
\usepackage{array}
\usepackage{multirow}
\usepackage[title]{appendix}
\usepackage{longtable}

\def \M {\mathcal{M}}

\def \Z {\mathbb{Z}}

\def \H {\mathcal{H}}

\begin{document}

\title{General Theory of Momentum-Space Nonsymmorphic Symmetry}

\author{Chen Zhang}
%\email[These authors contributed equally to this work.]{}
\affiliation{National Laboratory of Solid State Microstructures and Department of Physics, Nanjing University, Nanjing 210093, China}
\affiliation{The University of HongKong Shenzhen Institute of Research and Innovation, Shenzhen 518057, China}

\author{Z. Y. Chen}
%\email[These authors contributed equally to this work.]{}
\affiliation{National Laboratory of Solid State Microstructures and Department of Physics, Nanjing University, Nanjing 210093, China}
\affiliation{The University of HongKong Shenzhen Institute of Research and Innovation, Shenzhen 518057, China}

\author{Zheng Zhang}
%\email[These authors contributed equally to this work.]{}
\affiliation{National Laboratory of Solid State Microstructures and Department of Physics, Nanjing University, Nanjing 210093, China}
\affiliation{The University of HongKong Shenzhen Institute of Research and Innovation, Shenzhen 518057, China}

\author{Y. X. Zhao}
\email[]{yuxinphy@hku.hk}

\affiliation{Department of Physics and HKU-UCAS Joint Institute for Theoretical
	and Computational Physics at Hong Kong, The University of Hong Kong,
	Pokfulam Road, Hong Kong, China}
\affiliation{HK Institute of Quantum Science \& Technology, The University of Hong Kong,
	Pokfulam Road, Hong Kong, China}

\begin{abstract}
	As a fundamental concept of all crystals, space groups are partitioned into symmorphic groups and nonsymmorphic groups. Each nonsymmorphic group contains glide reflections or screw rotations with fractional lattice translations, which are absent in symmorphic groups. Although nonsymmorphic groups ubiquitously exist on real-space lattices, on the reciprocal lattices in momentum space, the ordinary theory only allows symmorphic groups.
	%since all symmetries leave the origin of the momentum space invariant and therefore contain no fractional translations of the reciprocal lattice. 
	In this work, we develop a novel theory for momentum-space nonsymmorphic space groups ($k$-NSGs), utilizing the projective representations of space groups. The theory is quite general: Given any $k$-NSGs in any dimensions, it can identify the real-space symmorphic space groups ($r$-SSGs) and construct the corresponding projective representation of the $r$-SSG that leads to the $k$-NSG.  To demonstrate the broad applicability of our theory, we show these projective representations and therefore all $k$-NSGs can be realized by gauge fluxes over real-space lattices.
	%For any given MSNG, the corresponding RSSG is identified as the one with the crystal class Fourier dual to that of the MSNG, and the multiplier for the projective representation is explicitly given. Nevertheless, all the projective representations here can be realized by tight-binding models with appropriate gauge fluxes, and therefore all MSNGs can be simulated by artificial crystals with their engineerable gauge fluxes.
	Our work fundamentally extends the framework of crystal symmetry, and therefore can accordingly extend any theory based on crystal symmetry, for instance the classification crystalline topological phases.
\end{abstract}

\maketitle

{\color{blue}\textit{Introduction}}
Space groups fundamentally characterize symmetries for all forms of natural crystals, including electronic materials, spin liquids and crystalline superconductors, as well as various artificial crystals, including photonic/acoustic crystals, cold atoms in optical lattices, periodic mechanical systems and electric-circuit arrays~\cite{bradley2010mathematical}.  Each crystal symmetry can be either nonsymmorphic or symmorphic, depending on whether it involves a fractional lattice translation. 
Two elementary examples are the glide reflection and screw rotation. The pure reflection or rotation changes the crystal, and the associated fractional lattice translation should be followed to leave the crystal invariant.  Hence, as a qualitative categorization, a space group is either symmorphic or nonsymmorphic groups, according to whether it contains nonsymmorphic symmetries. 

In three dimensions (two dimensions), there are $157$ ($4$) nonsymmorphic groups among $230$ ($17$) space groups. In fact, as the dimensionality increases, the proportion of nonsymmorphic groups becomes more and more dominant among all space groups~\cite{szczepanski2012geometry}. As we know, nonsymmorphic groups ubiquitously exist on \textit{real space} lattices. It has been a hot topic to discuss nontrivial (topological) properties arising from real-space nonsymmorphic space groups ($r$-NSGs)~\cite{Zak_nonsymmorphic,zhao2016nonsymmorphic,Sato_nonsymmorphic,Kane_nonsymmorphic}.

%Every space group $G$ is described by two groups: the translation group $\mathcal{L}$ as a normal subgroup, and the point group $P$ that is a finite subgroup of the $d$D orthogonal group $O(d)$. $P$ is isomorphic to the quotient group $G/\mathcal{L}$. All space groups can be categorized into two classes: symmorphic groups and nonsymmorphic groups. Each nonsymmorphic group contains at least one point-group element $R$ that not only properly or improperly rotates an arbitrary vector $\bm{r}$, but also translate $\bm{r}$ by a fraction $\bm{\tau}_R$ of the lattice translations~\cite{Point_Group_Notes}, i.e.,
%\begin{equation}\label{eq:tau_frac}
%	R:~\bm{r}\mapsto R\bm{r}+\bm{\tau}_R.
%\end{equation}
%In contrast, all point-group elements of a symmorphic group do not translate $\bm{r}$.

%The energy spectrum of a crystal is usually presented in momentum space as the band structure. This is enabled by the translation invariance of crystals, and the translation $\bm{t}\in \mathcal{L}$ is represented by $e^{i\bm{k}\cdot\bm{t}}$ with $\bm{k}$ the momentum. Then, apparently there is the translation symmetry in momentum space, described by the reciprocal lattice or momentum-space translation group $\widehat{\mathcal{L}}$, since $e^{i\bm{k}\cdot\bm{t}}=e^{i(\bm{k}+\bm{G})\cdot\bm{t}}$ for all $\bm{G}\in\widehat{\mathcal{L}}$. 
%This is the duality between real-space and momentum-space translation groups.

The situation is radically changed in \textit{momentum space} according to the ordinary theory of space groups~\cite{bradley2010mathematical}. The \textit{momentum space} is dual to the \textit{real space} under the Fourier transform. We have the reciprocal lattice in momentum space for each real-space lattice. And the momentum-space Hamiltonian $\mathcal{H}(\bm{k})$ is invariant under integral reciprocal-lattice translations, i.e.,   $\mathcal{H}(\bm{k}+\bm{K})=\mathcal{H}(\bm{k})$ for any reciprocal lattice vector $\bm{K}$. However, every symmetry operator only rotates $\bm{k}$ with NO fractional reciprocal-lattice translation~\cite{Point_Group_Notes}, and therefore no nonsymmorphic group exists in momentum space~\cite{bradley2010mathematical,SM}. 

Hence, a natural question arises: \textit{Is it possible to have nonsymmorphic groups in momentum space?}

%However, according to the conventional theory of space groups, there is no momentum-space nonsymmorphic symmetry. Any point group element $R$, no matter in a symmorphic or nonsymmorphic space group, merely rotates momenta $\bm{k}$~\cite{bradley2010mathematical,Nonsym_Notes}, i.e.,
%$R:~\bm{k}\mapsto R\bm{k}$.
%Hence, natural questions arise: how to realize nonsymmorphic symmetry in momentum space; whether we can develop a general theory for momentum-space nonsymmorphic groups (MNSGs).

%A notable recent advance is that a nonsymmorphic symmetry, namely the glide reflection symmetry, has been realized in momentum space of crystals through the projective symmetry algebra of real-space mirror reflections and translations. 

%However, through the projective representation, the momentum-space glide reflection not only inverses $k_x$ but also translates $k_y$ by $G_y/2$, with $G_y$ the reciprocal lattice constant of the $k_y$ direction. It can change the topology of momentum-space unit from the Brillouin zone (torus) to a Klein bottle. Accordingly, a novel topological invariant has been formulated over the Klein bottle, distinct from the Chern number over the Brillouin torus.

%As we have seen, the momentum-space nonsymmorphic symmetry is a characteristic feature for projective representations of space groups. However, to date only some models were presented. Hence, a general theory for nonsymmorphic groups in momentum space is badly demanded.

Here, we establish a novel theory of momentum-space nonsymmorphic space groups ($k$-NSGs) based on the \textit{projective} representations of space groups. Recall that a projective representation $\rho$ of a group $G$ is characterized by a multiplier $\nu$, i.e., 
\begin{equation}\label{eq:p_multi}
	\rho(g_1)\rho(g_2)=\nu(g_1,g_2)\rho(g_1g_2).
\end{equation} 
Here, $\rho(g)$ is the symmetry operator of $g\in G$, and $\nu(g_1,g_2)\in U(1)$ is a phase factor for any $g_1,g_2\in G$. It is appropriate multipliers that endow symmetry operators fractional reciprocal-lattice translations.

Our theory can exhaustively represent $k$-NSGs in any dimensions, i.e., for each $d$D $k$-NSG we can identify a real-space symmorphic space group ($r$-SSG) and the multiplier $\nu$ of the $r$-SSG that leads to the $k$-NSG. Particularly, we have explicitly tabulated the representations of all $157$ $3$D $k$-NSGs, and $4$ $2$D $k$-NSGs. 

All these multipliers of $r$-SSG and therefore all $k$-NSGs, can be realized by lattice models with gauge fluxes. This connects $k$-NSGs to spin liquids with emergent gauge fields~\cite{baskaran1993resonating,wen1999projective,Kitaev2006}, crystalline superconductors~\cite{Fan_Zhang_PRL}, and various artificial crystals with engineerable gauge fluxes~\cite{ozawa2019topological,ma2019topological,lu2014topological,yang2015topological,xue2021observation,imhof2018topolectrical,yu20204d,prodan2009topological,huber2016topological,cooper2019topological,dalibard2011colloquium}. 

%Among $230$ space groups, there are $157$ nonsymmorphic groups. As we shall prove in this Letter, all of the nonsymmorphic groups can be realized in momentum space by certain projective representations of the $73$ symmorphic groups in real space. Nevertheless, the procedure to construct the projective representation for any one of the $157$ MNSGs is explicitly given. 

%An exceptional case is recently pointed out~\cite{chen2022brillouin}. Under the projective representation of $2$D space group $Pm$, when the $y$-translation anti-commutes with $x$-reflection, the $x$-reflection acts nonsymmorphically in momentum space as the glide reflections i.e., $k_x$ is sent to $-k_x+G_y/2$. This is an evidence that the projective representation of space groups is the key point to realize MNSGs.

%A recent fascinating trend is to investigate novel topological phases protected by projective representations of space groups. 
%For instance, with time-reversal symmetry preserved, $\mathbb{Z}_2$ gauge fluxes can give rise to $\mathbb{Z}_2$-projective representations of crystal symmetries. The $\mathbb{Z}_2$ projective representations have been utilized for switching the spinless and spinful $PT$ symmetric topological classifications, and providing experimentally feasible proposals for realizing various topological phases. By breaking time-reversal symmetry, $U(1)$-projective representations can be realized by more generic gauge fluxes, which has been applied to the study of topological phases in the Hofstadter model.

%Clearly, the MNSGs substantially extend the paradigm of crystal symmetries. 
Under a $k$-NSG, the Hamiltonian is constrained by
\begin{equation}\label{eq:H-Constraint}
	U_R(\bm{k})\H(\bm{k})[U_R(\bm{k})]^\dagger=\mathcal{H}(R\bm{k}+\bm{\kappa}_R).
\end{equation}
Here, for each symmetry $R$, $\bm{\kappa}_R$ is the associated fractional reciprocal-lattice translation, and $U_R(\bm{k})$ is the unitary operator. 
%With the new symmetry constraints, we expect rich novel physics to be explored. Particularly, 
Since crystalline topological phases rely on symmetry constraints~\cite{fu2011topological,shiozaki2014topology,shiozaki2016topology,zhao2016nonsymmorphic}, and the novel constraints of Eq.~\eqref{eq:H-Constraint} can significantly extend the existing topological classifications~\cite{chiu2016classification,zhao2013topological,zhao2016unified}. %For instance, recently a novel Klein-bottle insulator has been proposed, which can also be understood in the present framework~\cite{chen2022brillouin}. 
Recently, it has been shown that projective symmetry can lead to fascinating topological  phases~\cite{chen2022brillouin,zhao2021switching,shao2021gauge,xue2022projectively,li2022acoustic,yang2022non,herzog2023hofstadter,meng2022spinful,chen2022brillouin,chen2023classification}. Hence, Eq.~\eqref{eq:H-Constraint} can open a broad avenue along this direction.

{\color{blue}\textit{An example: $k$-NSG $Pg$ from projective $Pm$}} Let us start with presenting a simple example to illustrate the main ideas before introducing our general theory.

Our example is the wallpaper group $Pm$. It is generated by translation symmetries $L_x$ and $L_y$ along the $x$ and $y$ directions, respectively, and the mirror reflection symmetry $M_x$ that inverses the $x$ coordinate. It is clear that in $Pm$
\begin{equation}
	M_xL_y=L_yM_x,
\end{equation}
i.e., $M_x$ and $L_y$ commute with each other since they act on different dimensions. 

We then consider a projective representation $\rho$ of $Pm$ with the multiplier $\nu$ satisfying
\begin{equation}\label{eq:LM_factor}
	\nu(M_x,L_y)=-1,\quad \nu(L_y,M_x)=1.
\end{equation}
Then, according to Eq.~\eqref{eq:p_multi}, $\rho(M_x)\rho(L_y)=-\rho(M_xL_y)$ and $\rho(L_y)\rho(M_x)=\rho(L_yM_x)$. Since $\rho(M_xL_y)=\rho(L_yM_x)$, we have the projective algebraic relation,
\begin{equation}\label{eq:p_LyMx}
	\rho(M_x)\rho(L_y)=-\rho(L_y)\rho(M_x).
\end{equation}

In momentum space, the translation operator $\rho(L_y)$ is decomposed into $\bm{k}$-components, $\rho_{\bm{k}}(L_y)=e^{ik_yb}$ with $b$ the lattice constant along the $y$ direction. Then, from \eqref{eq:p_LyMx}, we see $\rho(M_x)\rho_{\bm{k}}(L_y)[\rho(M_x)]^\dagger=-\rho_{\bm{k}}(L_y)$, i.e., $\rho(M_x)e^{ik_yb}[\rho(M_x)]^\dagger=-e^{ik_yb}=e^{i(k_y+G_y/2)b}$, where $G_y=2\pi/b$ is the reciprocal lattice constant along the $k_y$ direction. Hence, under the projective representation, $k_y$ is translated by a half of the reciprocal lattice constant, i.e.,
\begin{equation}
	\rho(M_x):~(k_x,k_y)\mapsto (-k_x,k_y+G_y/2).
\end{equation}
Here, the inversion of $k_x$ comes from the ordinary relation $\rho(M_x)\rho(L_x)=\rho(L_x^{-1})\rho(M_x)$.

From the above derivations, we see $\rho(M_x)$ acts on the momentum space as a nonsymmorphic symmetry, namely, a glide reflection, which stems from the multiplier $\nu$ specified by Eq.~\eqref{eq:LM_factor}.

Then, a natural question is how to realize the projective representation of Eq.~\eqref{eq:LM_factor} or \eqref{eq:p_LyMx}. It is well known that gauge fluxes can lead to projective representations. A general formulation is given in the Supplemental Materials~\cite{SM}. Specializing to Eq.~\eqref{eq:LM_factor}, a lattice model with appropriate gauge fluxes is illustrated in Fig.~\ref{fig:2D_model}(a).  

The Hamiltonian and the unitary operator $U_{M_x}$ for $M_x$ can be found in the Supplemental Material~\cite{SM}.
Importantly, the symmetry constraint is given by
\begin{equation}
	U_{M_x}\mathcal{H}(k_x,k_y)U^{\dagger}_{M_x}=\mathcal{H}(-k_x,k_y+G_y/2).
\end{equation}
One may write $\rho(M_x)$ in momentum space as $\rho(M_x)=U_{M_x}\mathcal{G}_{k_x}$. Here, $\mathcal{G}_{k_x}$ is the glide reflection in momentum space, with $\mathcal{G}_{k_x}(k_x,k_y)=(-k_x,k_y+G_y/2)$. The energy band structure is shown in Fig.~\ref{fig:2D_model}(b), and a constant energy cut is given in Fig.~\ref{fig:2D_model}(c). We observe that the band structure is indeed invariant under the glide reflection. Moreover, it is easy to verify Eq.~\eqref{eq:p_LyMx}, since $\rho(k_y)=e^{ik_yb}$.

In experiments, we need to realize the particular pattern of $\pi$ fluxes or negative hopping amplitudes of the model. A simple mechanism is illustrated in Fig.~\ref{fig:2D_model}(d). The low-energy effective hopping amplitude of two low-energy sites through a high-energy site is negative if the energy gap $\Delta$ is large enough. Thus, in principle, one can engineer negative hopping amplitudes by appropriately inserting high-energy sites [see Fig.~\ref{fig:2D_model}(d)]. More details and other mechanisms for realizing gauge fluxes can be found in the Supplemental Material~\cite{SM}.

\begin{figure}
	\centering
	\includegraphics[width=\columnwidth]{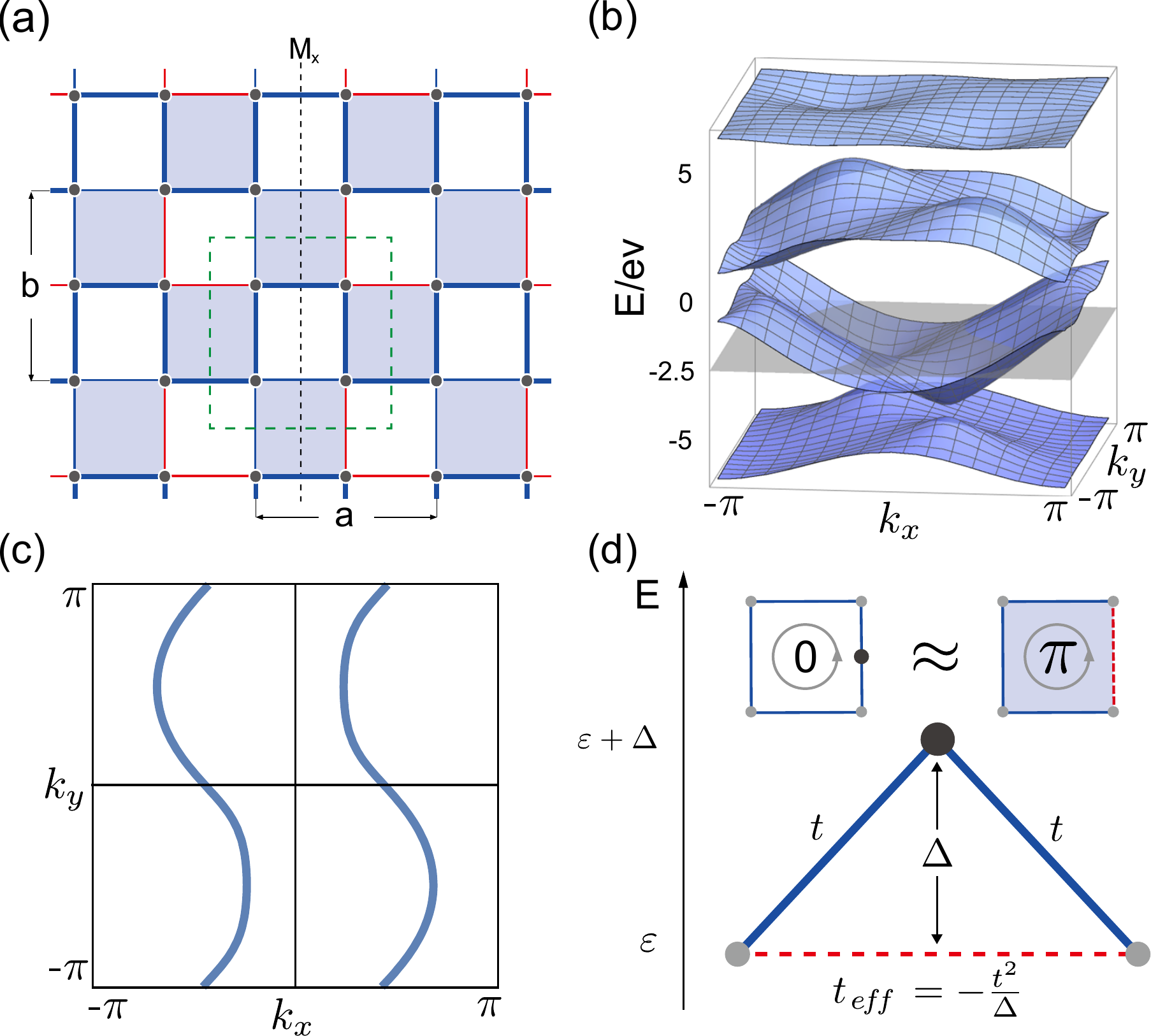}
	\caption{(a) The lattice model with the flux pattern preserving $Pm$. Blue and red bonds denote positive and negative hopping amplitudes, respectively. Each plaquette with $\pi$ flux is shadowed. (b) The band structure of the model. (c) The constant-energy contour of the band structure at $E=-0.25~\mathrm{eV}$. We observe that both (b) and (c) are invariant under the glide reflection $\mathcal{G}_x$. Note that $a=b=1$ is assumed in (b) and (c). (d) The effective negative hopping amplitude. Two (light) sites with energy $\epsilon$ hop to the (dark) site with energy $\epsilon+\Delta$ with amplitude $t>0$. The low-energy effective hopping amplitude between the two sites with energy $\epsilon$ is $-t^2/\Delta$. \label{fig:2D_model}}
\end{figure}

{\color{blue}\textit{Two questions}} We shall generalize the main ideas of the simple example into a general theory for all $k$-NSGs in any dimensions. Specifically, for each $k$-NSG, we shall answer two questions: 
(\textit{i}) What is the corresponding $r$-SSG? 
(\textit{ii}) With the $r$-SSG obtained, what is the multiplier that leads to the $k$-NSG?
	
To answer the two questions, we first look into how certain multipliers of projective representations can lead to fractional translations of momenta in the following two sections.

{\color{blue}\textit{The canonical multiplier}} To spell out the multipliers that we are interested in, we first introduce a basic formulation of $r$-SSGs. Each $r$-SSG contains two natural subgroups, namely the translation group $\mathcal{L}$ and the point group $P$. $\mathcal{L}$ consists of lattice translations with lengths $\bm{t}=\sum_i n^i \bm{e}_i$, where $n^i$ are integers, and $\bm{e}_i$ the primitive lattice vectors. Hence, the group elements can be written as
\begin{equation}\label{eq:Real_Lattice}
	\mathcal{L}=\{\bm{t}=\sum_i n^i \bm{e}_i,~ n^i\in \mathbb{Z}\}.
\end{equation} 
$\mathcal{L}$ can equally be interpreted as the collection of lattice sites related to the origin by lattice translations. Hereafter, we refer to $\mathcal{L}$ as the real-space translation group and lattice, interchangeably. The point group $P$ is a finite subgroup of the orthogonal group $O(d)$. $P$ compatibly operates on the lattice $\mathcal{L}$, i.e., $R\bm{t}\in \mathcal{L}$ for any $R\in P$ and $\bm{t}\in\mathcal{L}$. Such a compatible pair $(P,\mathcal{L})$ is referred to as a crystal class $D$. Accordingly, each $r$-SSG can be denoted by
$\mathcal{L}\rtimes_{D} P$, consisting of group elements $(\bm{t},R)$ with $\bm{t}\in \mathcal{L}$ and $R\in P$. The group multiplication is given by
\begin{equation}
	(\bm{t}_1,R_1)(\bm{t}_2,R_2)=(\bm{t}_1+R_1\bm{t}_2,R_1R_2).
\end{equation}
Notably, a point group may compatibly operate on more than one kinds of lattices, and therefore corresponds to a number of crystal classes. Consequently, in three dimensions, there are $32$ point groups, but $73$ crystal classes and therefore $73$ symmorphic space groups.

We are now ready to state one of key results, namely the multiplier, which is given by 
\begin{equation}\label{eq:the_muli}
\nu[(\bm{t}_1,R_1),(\bm{t}_2,R_2)]=e^{-i\bm{\kappa}_{R_1}\cdot R_1\bm{t}_2}.
\end{equation}
We shall show that $\bm{\kappa}_R$ is the fractional reciprocal-lattice translation associated to $R$, i.e., $R$ transforms $\bm{k}$ as
\begin{equation}\label{eq:Frac_R}
	R:~\bm{k}\mapsto R\bm{k}+\bm{\kappa}_{R}.
\end{equation}

Moreover, to make Eq.~\eqref{eq:the_muli} a multiplier , $\bm{\kappa}_R$ must satisfy the relation,
\begin{equation}\label{eq:kappa-cocycle}
	\bm{\kappa}_{R_1}+R_1\bm{\kappa}_{R_2}- \bm{\kappa}_{R_1R_2}\in \widehat{\mathcal{L}}.
\end{equation}
for any $R_1,R_2\in P$.
Here, $\widehat{\mathcal{L}}$ is the reciprocal lattice dual to $\mathcal{L}$. From $\bm{e}_i$, we can derive the primitive reciprocal-lattice vectors $\bm{G}_i$~\cite{Reciprocal_note}, and write $\widehat{\mathcal{L}}$ as
\begin{equation}
	\widehat{\mathcal{L}}=\{\bm{K}=\sum_i m^i \bm{G}_i,~m^i\in \mathbb{Z}\},
\end{equation} 
Note that  $e^{i\bm{t}\cdot \bm{K}}=1$ for any $\bm{t}\in \mathcal{L}$ and $\bm{K}\in\widehat{\mathcal{L}}$. Just like $\mathcal{L}$, $\widehat{\mathcal{L}}$ is referred to as the momentum-space translation group and reciprocal lattice interchangeably.

The rigorous derivations of Eqs.~\eqref{eq:the_muli} and \eqref{eq:kappa-cocycle} can be found in Appendix A, which are based on Mackey's canonical form of multipliers for semi-direct product groups~\cite{mackey1958unitary,mackey1989unitary}. 

{\color{blue}\textit{Fractional translations of momenta}} 
We then elucidate the meaning of $\bm{\kappa}_R$ in a projective representation $\rho$ with the multiplier of Eq.~\eqref{eq:the_muli}. It is significant to note that with the multiplier of Eq.~\eqref{eq:the_muli}, we can derive from Eq.~\eqref{eq:p_multi} the projective algebraic relation,
\begin{equation}\label{eq:Conjugation}
	\rho(\bm{t}',R)\rho(\bm{t},1)=e^{-i\bm{\kappa}_{R}\cdot R\bm{t}}\rho(R\bm{t},1)\rho(\bm{t}',R),
\end{equation}
Here, $(\bm{t},1)$ is an arbitrary element of the translation subgroup of the $r$-SSG $\mathcal{L}\rtimes_D P$~\cite{SM}.

Let us recall that each $\bm{k}$ corresponds to the irreducible representation $\rho_{\bm{k}}$ of the translation subgroup $\mathcal{L}$, with 
$\rho_{\bm{k}}(\bm{t},1)=e^{i\bm{k}\cdot \bm{t}}$. Then, the transformation of $\rho(\bm{t}',R)$ on $\bm{k}$ is given by $\rho_{\bm{k}}(\bm{t},1)\mapsto \rho(\bm{t}',R)\rho_{\bm{k}}(\bm{t},1)[\rho(\bm{t}',R)]^\dagger$. From Eq.~\eqref{eq:Conjugation}, we find that
\begin{equation}
	e^{i\bm{k}\cdot\bm{t}}\mapsto e^{-i\bm{\kappa}_R\cdot R\bm{t}}e^{i\bm{k}\cdot R\bm{t}}=e^{i(-R^T\bm{\kappa}_R+R^T\bm{k})\cdot \bm{t}}.
\end{equation}
Here, $R^T$ is the transpose of $R$ with $R^{-1}=R^T$. Note that $\bm{\kappa}_{R^T}+R^T\bm{\kappa}_{R}\in \widehat{\mathcal{L}}$, which comes from Eq.~\eqref{eq:kappa-cocycle} with $R_1=R^T$ and $R_2=R$. The transformation can be simplified to be 
\begin{equation}
	e^{i\bm{k}\cdot\bm{t}}\mapsto e^{i(R^T\bm{k}+\bm{\kappa}_{R^{T}})\cdot \bm{t}}.
\end{equation}
Thus, we conclude that each $R\in P$ operates on momentum space as Eq.~\eqref{eq:Frac_R}. Thus, we have proved that $\rho(\bm{t}',R)$ acts as a nonsymmorphic symmetry in momentum space. %since it contains the fractional reciprocal-lattice translation $\bm{\kappa}_R$. 

It is insightful to compare $k$-NSGs with $r$-NSGs. Recall that for a $r$-NSG with the point group $P$. Each $R\in P$ is associated with a fractional lattice translation $\bm{\tau}_R$, and we have the relation,
\begin{equation}\label{eq:tau_cocycle}
	\bm{\tau}_{R_1}+R_1\bm{\tau}_{R_2}-\bm{\tau}_{R_1R_2} \in  \mathcal{L}
\end{equation}
for any $R_1,R_2\in P$ \cite{bradley2010mathematical,szczepanski2012geometry,SM}. Here, $\mathcal{L}$ is the translation subgroup of the $r$-NSG. We observe that Eqs.~\eqref{eq:kappa-cocycle} and \eqref{eq:tau_cocycle} have exactly the same form, except that one is in momentum space and the other in real space. This further confirms that the projective representations of Eq.~\eqref{eq:the_muli} give rise to $k$-NSGs through Eq.~\eqref{eq:Frac_R}.

Remarkably, the duality between $k$-NSGs and projective representations of $r$-SSGs with Eq.~\eqref{eq:the_muli} can be rigorously established at the cohomological level, which is treated in Appendix B.

\begin{figure}
	\centering
	\includegraphics[width=\columnwidth]{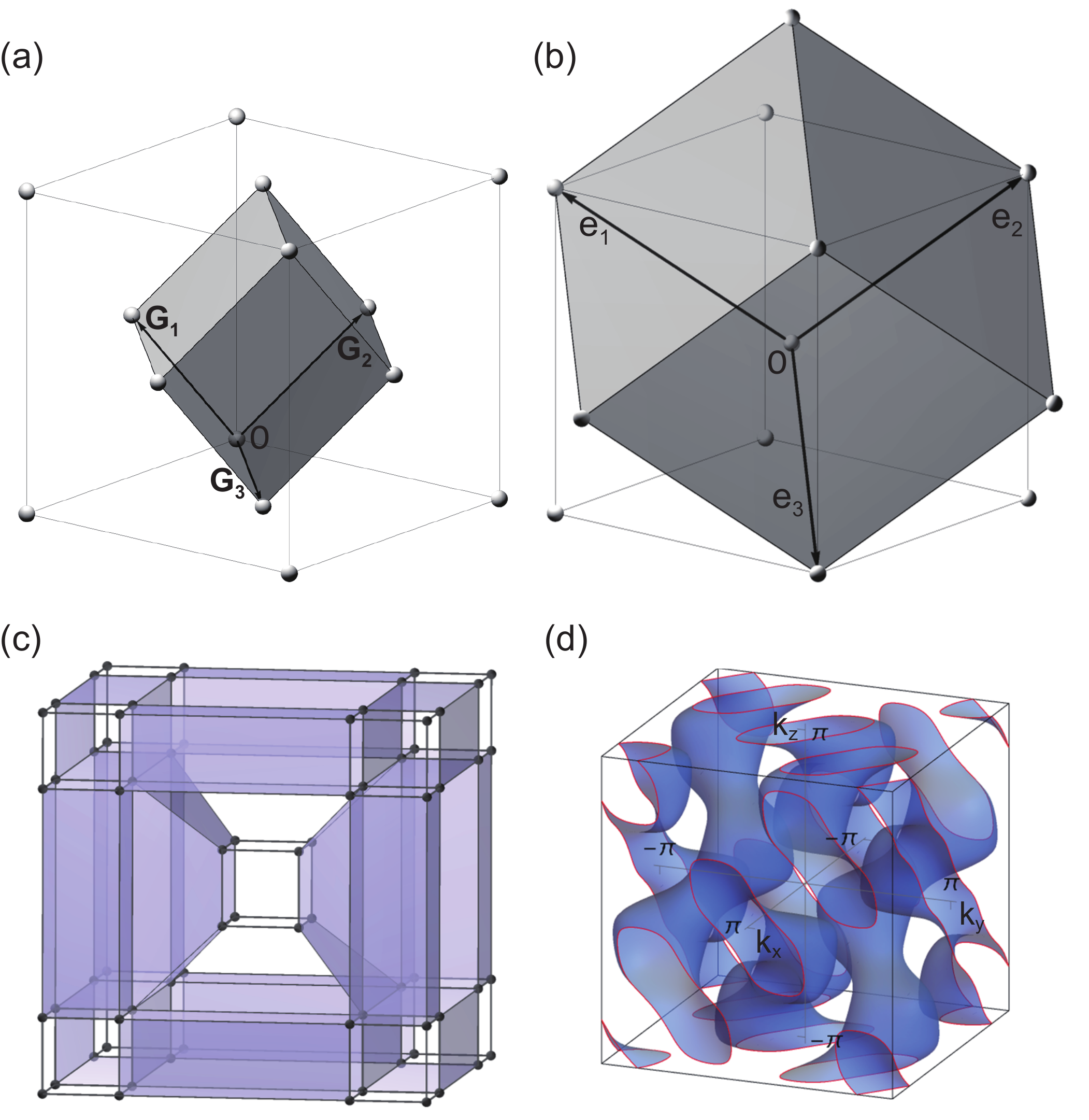}
	\caption{(a) and (b) are the dual face-centered and body-centered cubic lattices, respectively. The primitive translations are denoted by $\bm{G}_i$ and $\bm{e}_i$, and accordingly the fundamental domains are shadowed. The length of cube edge in (a) [(b)] is $2\pi$ (2). (c) The model for $k$-NSG $Fddd$. Each lattice site in (b) is substituted by a small cube, and $\pi$ fluxes are inserted through shadowed plaquettes connecting these small cubes. (d) A constant-energy contour of (c) preserving the $k$-NSG Fddd. See the Supplemental Material for more details~\cite{SM}. \label{fig:Dual_Lattices}}
\end{figure}

%\begin{figure}
%	\centering
%	\includegraphics[width=\columnwidth]{log.pdf}
%	\caption{(a) The algorithm for realizing MNSGs. Here, $Z^{1,\widehat{D}}(P,\mathbb{R}^d/\widehat{\mathcal{L}})$ consists of all solutions of \eqref{eq:kappa-cocycle}. (b) The duality of MNSGs and RNSGs from cohomology groups. \label{fig:Logics}}
%\end{figure}

{\color{blue}\textit{Answers and algorithm for all $k$-NSGs}} We are almost ready to answer the two questions that we initially proposed. One last piece needed is the concept called the Fourier duality between crystal classes, which is introduced below. 

By the Fourier transform of the real-space lattice $\mathcal{L}$, we obtain the reciprocal lattice $\widehat{\mathcal{L}}$. Both $\mathcal{L}$ and $\widehat{\mathcal{L}}$ are invariant under the same point group $P$. But, in terms of the $P$ actions, the crystal class $\widehat{D}$ of $\widehat{\mathcal{L}}$ may be different from $D$ of $\mathcal{L}$. $\widehat{D}$ is referred to as the Fourier dual of $D$. Since the Fourier transform is invertible, we have $\widehat{\widehat{D}}=D$, i.e., $D$ and $\widehat{D}$ are dual to each other, resembling the duality between $\mathcal{L}$ and $\widehat{\mathcal{L}}$. Hence, each crystal class is either self-dual or paired with its Fourier dual. For instance, in three dimensions, among the four crystal classes of point group $D_{2h}$, $mmmP$ and $mmmC$ are both self dual, while $mmmI$ and $mmmF$ are dual to each other. As illustrated in Fig.\ref{fig:Dual_Lattices}(a) and (b), the dual crystal classes $mmmF$ and $mmmI$ correspond to the face-centered cubic and body-centered cubic lattices, respectively.

For any given $k$-NSG with reciprocal lattice $\widehat{\mathcal{L}}$, point group $P$, and crystal class $\widehat{D}$,  it is now straightforward to answer the two questions. 
(\textit{i}) The $r$-SSG is $\mathcal{L}\rtimes_D P$, with $D$ dual to $\widehat{D}$ under the Fourier transform.
(\textit{ii}) To realize the $k$-NSG, the multiplier of the $r$-SSG, $\mathcal{L}\rtimes_D P$, is given by Eq.~\eqref{eq:the_muli}.

The answers lead to the following algorithm for constructing an arbitrary $k$-NSG. 

First, we write down the fractional translations $\bm{\kappa}_R$ satisfying Eq.~\eqref{eq:kappa-cocycle}. From standard textbooks, e.g., Ref.~\cite{bradley2010mathematical}, we can find the fractional translations $\bm{\tau}_R$ for the nonsymmorphic group. To get $\bm{\kappa}_R$, we just need to formally replace the real-space basis $\bm{e}_i$ by the reciprocal lattice basis $\bm{G}_i$. 

Second, from the reciprocal lattice $\widehat{\mathcal{L}}$ and the crystal class $\widehat{D}$ of the $k$-NSG, we determine the dual lattice $\mathcal{L}$ and dual class $D$. $\mathcal{L}$ and $D$ uniquely give the $r$-SSG $\mathcal{L}\rtimes_D P$ with $P$ the point group of the $k$-NSG.

Third, using Eq.~\eqref{eq:the_muli}, we write down the multiplier $\nu$ of $\mathcal{L}\rtimes_D P$ from $\bm{\kappa}_R$. The projective representation of $\mathcal{L}\rtimes_D P$ with multiplier $\nu$ gives the $k$-NSG.

{\color{blue}\textit{$2$D and $3$D $k$-NSGs}} Clearly, following the above algorithm, we can realize any $k$-NSG in any dimensions by constructing the $r$-SSG with the multiplier of Eq.~\eqref{eq:the_muli}. The projective representations for all $k$-NSGs in two and three dimensions can be found in the Supplemental Material~\cite{SM}. 

All the four $2$D nonsymmorphic space groups act on rectangular or square lattices, and therefore their crystal classes are all self dual. The case of two different crystal classes dual to each other occurs in three dimensions. Hence, we demonstrate our algorithm by such an example, namely $k$-NSG $Fddd$.

The lattice $\widehat{\mathcal{L}}$ of the MNSG $Fddd$ corresponds to the face-centered cube as illustrated in Fig.\ref{fig:Dual_Lattices}(b). The basis of $\widehat{\mathcal{L}}$ is given by 
\begin{equation*}
	\bm{G}_1=\pi(0,1,1),~\bm{G}_2=\pi(1,0,1),~\bm{G}_3=\pi(1,1,0).
\end{equation*}
The point group of $Fddd$ is $D_{2h}$, which is generated by three reflections $M_{k_i}$ with $i=1,2,3$. Here, $M_{k_{1,2,3}}$ inverse $k_{x,y,z}$, respectively. The three reflections are associated with fractional translations:
$\bm{\kappa}_{M_{k_i}}= \bm{G}_i/2$. The fractional translations for other elements of $D_{2h}$ can be derived from \eqref{eq:kappa-cocycle}. The crystal class of $Fddd$ is $mmmF$, which is dual to $mmmI$. Hence, the corresponding real-space symmorphic group is $Immm$ on the body-centered-cubic lattice $\mathcal{L}$ [see Figs.~\ref{fig:Dual_Lattices}(a) and (b)]. The basis of $\mathcal{L}$ is given by
\begin{equation}
	\bm{e}_1=(-1,1,1),~\bm{e}_2=(1,-1,1),~\bm{e}_3=(1,1,-1).
\end{equation}
The independent multiplier components are derived from Eq.~\eqref{eq:the_muli} as
\begin{equation}\label{eq:Immm_Multiplier}
	\nu((\bm{t}_1,M_{k_i}),(\bm{t}_2,R))=(-1)^{n_2^i},
\end{equation}
which corresponds to the projective representation $\rho$ satisfying
\begin{equation}
	\rho(M_{k_i})\rho(\bm{e}_i)\rho(M_{k_i})=-\rho(\bm{e}_1+\bm{e}_2+\bm{e}_3)
\end{equation}
with $i=1,2,3$. Other multiplier components and projective algebraic relations can be derived from them.

{\color{blue}\textit{Gauge-flux models for $k$-NSGs}}
Although not all projective representations of space groups can be realized by lattice models with gauge fluxes, the projective representations for $k$-NSGs can be realized in this way.  This can be inferred from the particular form of Eq.~\eqref{eq:the_muli}. It only modifies the algebraic relations between translations and point-group symmetries, namely, modifying the operation of the point group on the real-space lattice with additional phase factors [see Eq.~\eqref{eq:Conjugation}]. These phase factors can be realized by gauge fluxes, resembling the Aharonov–Bohm effect. See the Supplemental Material for a general formulation~\cite{SM}. %It is noteworthy that the different choice of gauge connections for the same flux configuration lead to equivalent projective representations and therefore to isomorphic MNSGs. 

A lattice model for $k$-NSGs $Fddd$ has been illustrated in Fig.\ref{fig:Dual_Lattices}(c). Each lattice site in Fig.\ref{fig:Dual_Lattices}(b) is substituted by a small cube without flux, and $\pi$ fluxes are inserted for shadowed plaquettes connecting these cubes. The lattice model realizes the projective representation of $Immm$ with the multiplier \eqref{eq:Immm_Multiplier}. A constant-energy contour of the model is shown in Fig.\ref{fig:Dual_Lattices}(d), which visualizes the $k$-NSG $Fddd$. The technical details for the lattice realization and all lattice realizations for the four $2$D $k$-NSGs can be found in the Supplemental Material \cite{SM}.

In fact, realizing negative hopping amplitudes is sufficient for all $2$D $k$-NSGs and 119 $3$D $k$-NSGs among the 157 ones in total. This is because these $k$-NSGs involve only half reciprocal-lattice translations, and therefore the multipliers are equal to $\pm 1$ according to Eq.~\eqref{eq:the_muli}. Thus, all $2$D and most $3$D $k$-NSGs can be readily realized by artificial crystals [see Fig.~\ref{fig:2D_model}(d)].

{\color{blue}\textit{Summary and discussions}}
In summary, we have revealed the intrinsic connection between projective representations and $k$-NSGs, based on Mackey's canonical form of multipliers for semi-direct product groups. From the connection, we can systematically construct any $k$-NSG by the projective representation of the corresponding $r$-SSG with the multiplier given by Eq.~\eqref{eq:the_muli}, which can be physically realized on the dual lattice with appropriate gauge fluxes. 

Our work substantially extends the scope of crystal symmetry, and deepens our understanding of the interplay between gauge structures and symmetry. We expect new avenues to be opened in topological physics and artificial crystals under the grand framework of $k$-NSGs.

\begin{acknowledgments}
	{\color{blue}\textit{Acknowledgements.}} This work is supported by National Natural Science Foundation of China (Grants No.~12161160315 and No.~12174181), Basic Research Program of Jiangsu Province (Grant No.~BK20211506), and the Guangdong-Hong Kong Joint Laboratory of Quantum Matter.
\end{acknowledgments}

\vskip 12pt
\appendix 

{\color{blue}\textit{Appendix A on canonical form of multipliers.}}
Mackey formulated a canonical form for multipliers of semi-direct product groups in the classic work \cite{mackey1958unitary}. The canonical form is introduced in the Supplemental Material with slightly changed conventions for our application~\cite{SM}. Since each $r$-SSG is a semi-direct product of the translation group $\mathcal{L}$ and the point group $P$, we can apply Mackey's canonical form to analyze the multipliers of $r$-SSGs.

Remarkably, according to Mackey's canonical form, any multiplier $\nu$ of an $r$-SSG can be decomposed as $\nu=\sigma\gamma\alpha$, with $\sigma$, $\alpha$ and $\gamma$ three elementary multipliers of the $r$-SSG. $\sigma$ and $\alpha$ are the restrictions of $\nu$ on the two subgroups $\mathcal{L}$ and $P$, respectively. Importantly, the multiplier of Eq.~\eqref{eq:the_muli} is in fact the $\gamma$ component, and any $\gamma$ component can be cast into the form of Eq.~\eqref{eq:the_muli}.  $\gamma$ is a multiplier that connects $\mathcal{L}$ and $P$. Hence, it is $\gamma$ that changes the algebraic relations between translation and point-group operators, and therefore leads to fractional reciprocal-lattice translations.

For our purpose, it is sufficient to presume that the restrictions of the $r$-SSG multiplier $\nu$ on the two subgroups $\mathcal{L}$ and $P$ are trivial, namely $\sigma=\alpha=1$. Then, in accord with Mackey's canonical form, the multiplier can be written as
\begin{equation}\label{eq:multiplier_a}
	\nu[(\bm{t}_1,R_1),(\bm{t}_2,R_2)]=\gamma(R_1\bm{t}_2,R_1).
\end{equation}
Here, $\gamma(\bm{t},R)$ is valued in $U(1)$, and satisfies the following conditions: 
\begin{equation}\label{eq:linearity}
	\gamma(\bm{t}_1+\bm{t}_2,R)=\gamma(\bm{t}_1,R)\gamma(\bm{t}_2,R),
\end{equation}
and
\begin{equation}\label{eq:1-cocycle}
	\gamma(\bm{t},R_1R_2)=\gamma(\bm{t},R_1)\gamma(R_1^{T}\bm{t},R_2).
\end{equation}
It is straightforward to check that the two conditions Eqs.~\eqref{eq:multiplier_a} and \eqref{eq:1-cocycle} are sufficient for making $\nu$ a multiplier, i.e., $\nu(g_1,g_2)\nu(g_1g_2,g_3)=\nu(g_1,g_2g_3)\nu(g_2,g_3)$ for all $g_1,g_2,g_3\in \mathcal{L}\rtimes_{D}P$.

The first equation implies $\gamma(*,R)$ with $R$ fixed is a homomorphism from $\mathcal{L}$ to $U(1)$, and therefore $\gamma$ takes the form:
\begin{equation}\label{eq:gamma}
	\gamma(\bm{t},R)=e^{-i\bm{\kappa}_R\cdot \bm{t}},
\end{equation}
where $\bm{\kappa}_R$ specifies the homomorphism for each $R\in P$. Substituting Eq.~\eqref{eq:gamma} into Eq.~\eqref{eq:multiplier_a}, we obtain Eq.~\eqref{eq:the_muli}. Substituting Eq.~\eqref{eq:gamma} into Eq.~\eqref{eq:1-cocycle} gives the significant relation in Eq.~\eqref{eq:kappa-cocycle}.

{\color{blue}\textit{Appendix B on cohomological equivalence.}}
In fact, the duality between $k$-NSGs and projective representations of $r$-SSGs with Eq.~\eqref{eq:the_muli} can be solidly established at the cohomological level. 

Let us recall that to identify a $r$-NSG, it is not sufficient to only specify the fractional translation $\bm{\tau}_R$  for each point-group element $R$. This is because for an arbitrary vector $\bar{\bm{r}}$, we can construct the fractional translations,
\begin{equation}
	\delta\bar{\bm{r}}_R=R\bar{\bm{r}}-\bar{\bm{r}},
\end{equation}
which obviously satisfy Eq.~\eqref{eq:tau_cocycle}. But such fractional translations are trivial, since they correspond to essentially symmorphic space groups~\cite{szczepanski2012geometry}. That is, $\delta\bar{\bm{r}}_R$ comes from the displacement $\bar{\bm{r}}$ of the point-group reference point and the coordinate origin. Moreover, $\bm{\tau}_R$ and $\tilde{\bm{\tau}}_R$ are equivalent if their difference $\tilde{\bm{\tau}}_R-\bm{\tau}_R=\delta\bar{\bm{r}}_R$ for some $\bar{\bm{r}}$, since they can be equated by shifting the coordinate origin by $\bar{\bm{r}}$.

Hence, to fully establish the concept of $k$-NSGs, we need to check whether such momentum-space fractional translations,
\begin{equation}
	\delta\bm{\bar{\bm{k}}}_R=R\bar{\bm{k}}-\bar{\bm{k}},
\end{equation}
correspond to trivial multipliers.
Here, $\bar{\bm{k}}$ is an arbitrary vector in momentum space. To see this, it is noticed that the corresponding multiplier is given by
\begin{equation}
	\nu_{\bar{\bm{k}}}[(\bm{t}_1,R_1),(\bm{t}_2,R_2)]=e^{i\bar{\bm{k}}\cdot R_1 \bm{t}_2}e^{-i\bar{\bm{k}}\cdot\bm{t}_2}.
\end{equation}
The multiplier is trivial, because it can be induced from an ordinary representation by multiplying each operator $\rho(\bm{t},R)$ with the phase factor $\chi(\bm{t},R)=e^{i\bar{\bm{k}}\cdot\bm{t}}$. That is, it is equal to $\{\chi[(\bm{t}_1,R_1)(\bm{t}_2,R_2)]/\chi(\bm{t}_1,R_1)\chi (\bm{t}_2,R_2)\}$. Hence, we have confirmed that if $\tilde{\bm{\kappa}}_R-\bm{\kappa}_R=\delta\bar{\bm{k}}_R$ for some $\bar{\bm{k}}$, $\tilde{\bm{\kappa}}_R$ and $\bm{\kappa}_R$ are equivalent, since they correspond to equivalent projective representations.

Thus, given a point group $P$, all possibilities of $k$-NSGs are characterized by equivalence classes of solutions of Eq.~\eqref{eq:kappa-cocycle}. This is just the `twisted' first cohomology group $H^{1,\widehat{D}}(P,\mathbb{R}^d/\widehat{\mathcal{L}})$.
Here, $\mathbb{R}^d/\widehat{\mathcal{L}}$ denotes the fundamental domain in momentum space, namely the Brillouin zone, in which $\bm{\kappa}_R$ is valued. In parallel, all possible real-space nonsymmorphic groups are given by $H^{1,D}(P,\mathbb{R}^d/\mathcal{L})$. Starting with $(P,D)$, we can first work out $H^{1,\widehat{D}}(P,\mathbb{R}^d/\widehat{\mathcal{L}})$ and  $H^{1,D}(P,\mathbb{R}^d/\mathcal{L})$, and then exhaust all the $k$-NSGs and $r$-NSGs, respectively.

\bibliographystyle{apsrev4-1}
\bibliography{references}

\begin{thebibliography}{46}
\bibitem [{Poi()}]{Point_Group_Notes}%
  \BibitemOpen
  \href@noop {} {}\bibinfo {note} {An element $R$ of $O(d)$ is called a proper
  or improper rotation according to $\mathrm{Det}(R)=\pm 1$.}\BibitemShut
  {Stop}%
\bibitem [{Rec()}]{Reciprocal_note}%
  \BibitemOpen
  \href@noop {} {}\bibinfo {note} {In three dimensions, the primitive
  reciprocal lattice vectors are given by
  $\bm{G}_i=2\pi\epsilon_{ijk}\bm{e}_j\times\bm{e}_k/|\bm{e}_1\cdot(\bm{e}_2\times\bm{e}_3)|$.}\BibitemShut
  {Stop}%
\end{thebibliography}

\onecolumngrid
\renewcommand{\theequation}{S\arabic{equation}}
\setcounter{equation}{0}
\renewcommand{\thefigure}{S\arabic{figure}}
\setcounter{figure}{0}
\newpage
\section{Supplemental Materials for \\``General Theory of Momentum-Space Nonsymmorphic Groups"}

\section{ The canonical form of multipliers} 

Here, we provide a brief introduction to Mackey's canonical form of multipliers for semidirect product groups.

A projective representation $\rho$ of a group $G$ is characterized by a multiplier $\nu$, defined as 
\begin{equation}
\rho(g_1)\rho(g_2)=\nu(g_1,g_2)\rho(g_1g_2),
\end{equation}
where $\nu(g_1,g_2)\in U(1)$ for all $g_1,g_2\in G$. We refer to $\rho$ as a $\nu$-representation of $G$. The associativity condition $[\rho(g_1)\rho(g_2)]\rho(g_3)=\rho(g_1)[\rho(g_2)\rho(g_3)]$ requires the multiplier $\nu$ to satisfy the self-consistency equation
\begin{equation}\label{cocyleeq}
\nu(g_1,g_2)\nu(g_1g_2,g_3)=\nu(g_1,g_2g_3)\nu(g_2,g_3), \ \ \ \ \forall g_1,g_2,g_3 \in G.
\end{equation}
If we transform $\rho(g)$ to $\rho^{\prime}(g)=\chi(g)\rho(g)$ using a phase factor $\chi(g)\in U(1)$, then the multiplier $\nu(g_1,g_2)$ is transformed as $\nu^{\prime}(g_1,g_2)=\nu(g_1,g_2)\chi(g_1)\chi(g_2)/\chi(g_1g_2)$. A function of the form $\chi(g_1)\chi(g_2)/\chi(g_1g_2)$ is known as a coboundary. Two multipliers that differ from each other by a coboundary are considered equivalent.

If group $G$ is a semidirect product of $N$ and $H$, $G=N\rtimes H$, then every $g\in G$ can be uniquely written as $g=nh$ with $n\in N$ and $h\in H$. Here, $N$ is a normal subgroup of $G$, i.e., for all $g\in G, n\in N$, the conjugation $c_{g}n:=gng^{-1} \in N$.

Any multiplier $\nu$ of $G=N\rtimes H$ can be decomposed as \cite{mackey1958unitary}
\begin{equation}\label{eq:decomposition}
\nu(n_1h_1, n_2h_2)=\sigma(n_1,c_{h_1}n_2)\gamma(c_{h_1}n_2,h_1)\alpha(h_1,h_2)
\end{equation}
up to a coboundary $\chi$. Here, $\sigma$ and $\alpha$ are the restrictions of $\nu$ to the subgroups $N$ and $H$ respectively. $\gamma$ is a $U(1)$-valued function, which can be interpreted as the ``gluing'' part. Equation~\eqref{eq:decomposition} is called Mackey's canonical form of multipliers.
It is worth noting that, for our purpose the middle term $\gamma$ has been expressed in a different but equivalent form compared to the original form presented in Mackey's paper~\cite{mackey1958unitary}. The validity of the canonical form will be demonstrated in the following. 

\begin{proof} 
	Let $\nu^{\prime}$ be a multiplier for $G$ and $\rho^{\prime}$ be any $\nu^{\prime}$-representation. Then $\nu^{\prime}(n,h)\rho^{\prime}(nh)=\rho^{\prime}(n)\rho^{\prime}(h)$ for all $n\in N,h\in H$. Let $\rho(nh)=\nu^{\prime}(n,h)\rho^{\prime}(nh)$. Then $\rho$ is a $\nu$-representation for a multiplier $\nu$ which is similar to $\nu^{\prime}$. Since $\nu^{\prime}(n,E)=\nu^{\prime}(E,h)=1$ for all $n,h\in G$, it follows that $\rho(nh)=\rho^{\prime}(n)\rho^{\prime}(h)=\rho(n)\rho(h)=A(n)B(h)$ where $A$ and $B$ denote the restrictions of $\rho$ to $N$ and $H$ respectively. Moreover $A$ is a $\sigma$-representation of $N$ and $B$ is an $\alpha$-representation of $H$ where $\sigma$ and $\alpha$ denote the restrictions of $\nu$ to $N$ and $H$ respectively. Now 
	\begin{equation}
	\rho(n_1h_1n_2h_2)=\rho(n_1h_1n_2h_1^{-1}h_1h_2)=A(n_1c_{h_1}n_2)B(h_1h_2). 
	\end{equation}
	Thus 
	\begin{equation}
	\sigma(n_1,c_{h_1}n_2)\alpha(h_1,h_2)A(n_1)B(h_1)A(n_2)B(h_2)=\nu(n_1h_1,n_2h_2)A(n_1)A(c_{h_1}n_2)B(h_1)B(h_2),
	\end{equation}
	and then 
	\begin{equation}
	\nu(n_1h_1,n_2h_2)/[\sigma(n_1,c_{h_1}n_2)\alpha(h_1,h_2)]=A(c_{h_1}n_2)^{-1}B(h_1)A(n_2)B(h_1)^{-1}.
	\end{equation}
	The right hand of the equation depends only on $n_2$ and $h_1$. Denoting it by $\gamma(c_{h_1}n_2,h_1)$, we obtain Eq.~\eqref{eq:decomposition}.
\end{proof}

With this decomposition, the self-consistency condition Eq. (\ref{cocyleeq}) can be satisfied by requiring $\sigma,\gamma,\alpha$ to satisfy 
\begin{equation}\label{co2}
\begin{aligned}
\sigma(n_1,n_2)\sigma(n_1n_2,n_3)=\sigma(n_1,n_2n_3)\sigma(n_2,n_3), \ \ \ \ \forall n_1,n_2,n_3 \in N,\\
\alpha(h_1,h_2)\alpha(h_1h_2,h_3)=\alpha(h_1,h_2h_3)\alpha(h_2,h_3), \ \ \ \ \forall h_1,h_2,h_3 \in H,
\end{aligned}
\end{equation}
and 
\begin{equation}\label{eq:Consistency_a}
\frac{\sigma(c_{h^{-1}}n_1,c_{h^{-1}}n_2)}{\sigma(n_1,n_2)}=\frac{\gamma(n_1n_2,h)}{\gamma(n_1,h)\gamma(n_2,h)},
\end{equation}
\begin{equation}\label{eq:Consistency_b}
\gamma(n,h_1h_2)=\gamma(n,h_1)\gamma(c_{h_1^{-1}}n,h_2).
\end{equation}
The derivation can be found in \cite{mackey1958unitary}. Here, we also provide a simple proof.
\begin{proof}
	Equation~\eqref{co2} is automatically satisfied since $\sigma$ and $\alpha$ are multipliers of projective representation A and B respectively. And now the condition \eqref{cocyleeq} for $\nu$ to be a multiplier of $G$ requires that $\gamma$ and $\sigma$ satisfy the following identity:
	\begin{equation}\label{eq:multiplier}
	\begin{split}
	\gamma(c_{h_1h_2}n_3,h_1h_2)\gamma(c_{h_1}n_2,h_1)\sigma(n_1c_{h_1}n_2,&c_{h_1h_2}n_3)\sigma(n_1,c_{h_1}n_2)\\
	&=\gamma[c_{h_1}(n_2c_{h_2}n_3),h_1]\gamma(c_{h_2}n_3,h_2)\sigma[n_1,(c_{h_1}n_2)c_{h_1h_2}n_3]\sigma(n_2,c_{h_2}n_3).
	\end{split}
	\end{equation}
	Setting $h_2=E$ and using Eq.~\eqref{co2} to reduce this equation, we get 
	\begin{equation}
	\gamma(c_{h_1}n_3,h_1)\gamma(c_{h_1}n_2,h_1)\sigma(c_{h_1}n_2,c_{h_1}n_3)=\sigma(n_2,n_3)\gamma[c_{h_1}(n_2n_3),h_1].
	\end{equation}
	Let $h_1=h$ and $n_2=c_{h^{-1}}n_1,n_3=c_{h^{-1}}n_2$. Then we get Eq. \eqref{eq:Consistency_a}. Using the Eq.~\eqref{eq:Consistency_a} to simplify Eq.~\eqref{eq:multiplier} we get Eq.~\eqref{eq:Consistency_b}. Thus Eq.~\eqref{eq:multiplier} is equivalent to Eq.~\eqref{eq:Consistency_a} and Eq.~\eqref{eq:Consistency_b}.
\end{proof}

Equation~\eqref{co2} represents the self-consistency equations for multipliers of subgroups $N$ and $H$, while Eq.~\eqref{eq:Consistency_a} and Eq.~\eqref{eq:Consistency_b} denote additional self-consistency conditions for  $\sigma$ and $\gamma$. 

In the case where $G$ is a symmorphic space group, it can be expressed as $G=\mathcal{L}\rtimes_D P$, where $\mathcal{L}$ is the translation group and $P$ is the point group. In this scenario, $\sigma$ and $\alpha$ correspond to the multipliers of $\mathcal{L}$ and $P$, respectively. We can represent elements in group $G$ as $(\bm{t},R)$, with $\bm{t}\in\mathcal{L}$ and $R\in P$.

Assuming that both $\sigma$ and $\alpha$ are trivial, any multiplier $\nu$ of $G$ can be written as:
\begin{equation}
\nu[(\bm{t}_1,R_1),(\bm{t}_2,R_2)]=\gamma(R_1\bm{t}_2,R_1),
\end{equation}
where $c_{R_1}\bm{t}_2$ is denoted as $R_1\bm{t}_2$ for brevity.

\section{Momentum-space nonsymmorphic space groups}

We demonstrate in the main text that all $k$-space nonsymmorphic space groups ($k$-NSGs) of reciprocal crystal class $\widehat{D}$ can be obtained from real-space symmorphic groups of crystal class $D$ with suitable multipliers. Here, we provide a comprehensive list of the four $k$-NSGs in 2D and the complete set of 157 $k$-NSGs in 3D.

\subsection{2D $k$-NSGs}
The Tab.\ref{tab:2D_k-NSGs} shows the realizations of the 4 2D $k$-NSGs. The four 2D nonsymmorphic groups act on rectangular or square lattices, and therefore their crystal classes are self dual. We may choose $\bm{G}_x=2\pi(1,0)$ and $\bm{G}_y=2\pi(0,1)$ as the basis for the reciprocal lattice $\widehat{\mathcal{L}}$, and the corresponding basis for real-space lattice is given by $\bm{e}_x=(1,0)$ and $\bm{e}_y=(0,1)$. 
\begin{table*}[!ht]
	\centering
	\begin{tabular}{c|c|c|c|c|c|c}
		\hline 
		$P$&Presentation&$k$-NSG& Fractional translation&$r$-SSG&Multiplier component &Projective algebra \\
		\hline \hline
		$D_1$&$M_x|M_x^2$&$Pg$ &$\bm{\kappa}_{M_x}=\bm{G}_y/2$& $Pm$ &$\nu[(\bm{t}_1,M_x),(\bm{t}_2,R)]=(-1)^{n_2^y}$ &$\{\rho(e_y),\rho(M_x)\}=0$\\
		\hline
		\multirow{3}{*}{$D_2$}&\multirow{3}{*}{$\begin{array}{c} M_x,M_y|\\M_x^2,M_y^2,(M_xM_y)^2\end{array}$}&$Pmg$ & $\bm{\kappa}_{M_x}=\bm{G}_y/2$ &\multirow{3}{*}{$Pmm$} &$\begin{array}{c}\nu[(\bm{t}_1,M_xM_y),(\bm{t}_2,R)]=(-1)^{n_2^y},\\\nu[(\bm{t}_1,M_x),(\bm{t}_2,R)]=(-1)^{n_2^y}\end{array}$&$\begin{array}{c}\{\rho(e_y),\rho(M_x)\}=0,\\\{\rho(e_y),\rho(M_xM_y)\}=0\end{array}$\\
		\cline{3-4}\cline{6-7}
		&&$Pgg$ &$\begin{array}{c}\bm{\kappa}_{M_x}=\bm{G}_y/2\\\bm{\kappa}_{M_y}=\bm{G}_x/2\end{array}$  &&$\begin{array}{c}\nu[(\bm{t}_1,M_xM_y),(\bm{t}_2,R)]=(-1)^{n_2^x+n_2^y},\\\nu[(\bm{t}_1,M_{i}),(\bm{t}_2,R)]=(-1)^{n_2^{\bar{i}}}\end{array}$&$\begin{array}{c}\{\rho(e_y),\rho(M_x)\}=0,\\\{\rho(e_x),\rho(M_y)\}=0\end{array}$ \\
		\hline 
		$D_4$&$\begin{array}{c} C,M_x|\\C^4,M_x^2,(CM_x)^2\end{array}$&$P4g$&$\bm{\kappa}_{M_x}=(\bm{G}_x+\bm{G}_y)/2 $  &$P4m$ &$\nu[(\bm{t}_1,M_x),(\bm{t}_2,R)]=(-1)^{n_2^x+n_2^y}$&$\begin{array}{c}[\rho(e_y)\rho(M_x)]^2=-1,\\\{\rho(e_y),\rho(M_x)\}=0\end{array}$\\
		\hline
	\end{tabular}
	\caption{Momentum-space nonsymmorphic groups in two dimensions. The first column $P$ lists the point groups. The second presents the generators and algebraic relations (separated by `$|$') for each point group. Note that all terms after `$|$' equal the identity $E$. The $k$-NSGs are listed in the third column, and the corresponding fractional translations for generators are given in the fourth column. `$r$-SSG' stands for the real-space symmorphic group for each $k$-NSG. The nontrivial multiplier components and independent projective algebraic relations are, respectively, given in the last two columns. \label{tab:2D_k-NSGs}}
\end{table*}

We now exemplify our results in Tab.\ref{tab:2D_k-NSGs} by $k$-NSG $Pg$. The point group is $D_1=\{E,M_x\}$, with $M_x$ inversing $k_x$. The fractional translation for $M_x$ is given by $\bm{\kappa}_{M_x}=\bm{G}_y/2$, i.e., $M_x:\bm{k}\rightarrow(-k_x,k_y+\bm{G}_y/2)$. The corresponding $r$-SSG is $Pm$ in the same crystal class as $Pg$.  According to the main text, the multiplier of $Pm$ is given by 
\begin{equation}
\nu[(\bm{t_1},E),(\bm{t}_2,R)]=1,\quad \nu[(\bm{t}_1,M_x),(\bm{t}_2,R)]=(-1)^{n_2^y},
\end{equation}
where $R=E$ or $M_x$ and $\bm{t}_i=n_i^x\bm{e}_x+n_i^y\bm{e}_y$. With this multiplier, it is straightforward to check the algebraic relation $\{\rho(\bm{e}_y),\rho(M_x)\}=0$ for the projective representation $\rho$ with $\nu$. In the next section, we will show the model realization of these projective algebras. And the table will be further explained.

\subsection{3D $k$-NSGs}
We present a comprehensive list of all 157 $k$-space nonsymmorphic space groups ($k$-NSGs) in 3D, including their reciprocal classes, associated $r$-space symmorphic space groups ($r$-SSGs), and crystal classes.

The following table provides the crystal system and point group of each $k$-NSG in the first and second columns, respectively. The third column displays the $k$-NSGs that can be achieved through the projective representations of the same real-space symmorphic group listed in the last column. The fourth column indicates the reciprocal arithmetic crystal classes of the $k$-NSGs, while the fifth column corresponds to the dual real-space arithmetic crystal class. Lastly, the last column presents the $r$-SSGs necessary for realizing each $k$-NSG.
\begin{center}
    {
    \setlength{\tabcolsep}{1.1mm}
    \renewcommand\arraystretch{1.2}
    \begin{longtable}{|c|c|c|c|c|c|}
    \caption{All 3D $k$-NSGs and corresponding $r$-SSGs.}\\
    \hline 
    Crystal system & Point group & \makebox[0.35\textwidth][c]{$k$-NSGs} & Reciprocal class &Crystal class&$r$-SSG\\
    \hline
    \endfirsthead

    \multicolumn{6}{c}{{\bfseries\tablename\ \thetable{}--continued from previous page}}\\
    \hline 
    Crystal system & Point group & \makebox[0.35\textwidth][c]{$k$-NSGs} & Reciprocal class &Crystal class&$r$-SSG\\
    \hline
    \endhead 

    \hline\multicolumn{6}{l}{{Continued on next page}}\\
    \endfoot

    \hline\endlastfoot

    \multirow{5}{*}{Monoclinic}& $C_2$ & $P2_1$ & $2P$ & $2P$ &$P2$\\
        \cline{2-6}
        & \multirow{2}{*}{$C_s$} & $Pc$ & $mP$ & $mP$&$Pm$\\
        \cline{3-6}
        & & $Cc$ & $mC$ & $mC$&$Cm$\\
        \cline{2-6}
        &\multirow{2}{*}{$C_{2h}$} &$P2_1/m,P2/c,P2_1/c$ & $2/mP$ &$2/mP$ &$P2/m$\\
        \cline{3-6}
        & &$C2/c$ & $2/mC$ & $2/mC$&$C2/m$\\
        \hline
         & \multirow{3}{*}{$D_2$} & $P222_1,P2_12_12,P2_12_12_1$ &$222P$ & $222P$&$P222$\\
        \cline{3-6}
        && $C222_1$ &$222C$ & $222C$&$C222$\\
        \cline{3-6}
        && $I2_12_12_1$ & $222I$ & $222F$&$F222$\\
        \cline{2-6}
        \multirow{12}{*}{Orthorhombic}&\multirow{6}{*}{$C_{2v}$}& $\begin{array}{c}Pmc2_1,Pcc2,Pma2,Pca2_1,\\Pnc2,Pmn2_1,Pba2,Pna2_1,Pnn2\end{array}$ & $mm2P$ & $mm2P$&$Pmm2$\\
        \cline{3-6}
        && $Cmc2_1,Ccc2$ & $mm2C$ & $mm2C$&$Cmm2$\\
        \cline{3-6}
        &&$Aem2,Ama2,Aea2$ & $mm2A$ & $mm2A$&$Amm2$\\
        \cline{3-6}
        && $Ibam,Ibca,Imma$ & $mm2I$ & $mm2F$&$Fmm2$\\
        \cline{3-6}
        && $Fdd2$ & $mm2F$ &$mm2I$&$Imm2$\\
        \cline{2-6}
        &\multirow{4}{*}{$D_{2h}$}& $\begin{array}{c}Pnnn,Pccm,Pban,Pmma,Pnna,\\Pmna,Pcca,Pbam Pccn,Pbcm,\\Pnnm,Pmmn,Pbcn,Pbca,Pnma\end{array}$ & $mmmP$ & $mmmP$&$Pmmm$\\
        \cline{3-6}
        & & $Cmce,Cmcm,Cccm,Cmme,Ccce$ & $mmmC$ & $mmmC$&$Cmmm$\\
        \cline{3-6}
        & & $Ibam,Ibca,Imma$ & $mmmI$ & $mmmF$&$Fmmm$\\
        \cline{3-6}
        && $Fddd$ & $mmmF$ & $mmmI$&$Immm$\\
        \hline
        \multirow{19}{*}{Tetragonal}&\multirow{2}{*}{$C_4$} & $P4_1,P4_2,P4_3$ & $4P$ &$4P$&$P4$\\
        \cline{3-6}
        && $I4_1$ & $4I$ & $4I$&$I4$\\
        \cline{2-6}
        &\multirow{2}{*}{$C_{4h}$}&$P4_2/m,P4/n,P4_2/n$ & $4/mP$ &$4/mP$&$P4/m$\\
        \cline{3-6}
        && $I4_1/\text{a}$ & $4/mI$ & $4/mI$&$I4/m$\\
        \cline{2-6}
        &\multirow{2}{*}{$D_4$}& $\begin{array}{c}P42_12,P4_122,P4_12_12,\\P4_222,P4_22_12,P4_322,P4_32_12\end{array}$ & $422P$ & $422P$&$P422$\\
        \cline{3-6}
        && $I4_122$ & $422I$ &$422I$&$I422$\\ 
        \cline{2-6}
        &\multirow{2}{*}{$C_{4v}$}& $\begin{array}{c}P4nc,P4_2mc,P4_2bc,\\ P4bm,P4_2cm,P4_2nm,P4cc \end{array}$ & $4mmP$ & $4mmP$&$P4mm$\\
        \cline{3-6}
        && $I4cm,I4_1md,I4_1cd$ & $4mmI$ & $4mmI$&$I4mm$\\
        \cline{2-6}
        &\multirow{4}{*}{$D_{2d}$}& $P\bar{4}2c,P\bar{4}2_1m,P\bar{4}2_1c$& $\bar{4}2mP$ & $\bar{4}2mP$ &$P\bar{4}2m$\\
        \cline{3-6}
        && $P\bar{4}c2,P\bar{4}bc,P\bar{4}n2$ & $\bar{4}m2P$ & $\bar{4}m2P$&$P\bar{4}m2$\\
        \cline{3-6}
        && $I\bar{4}2d$ & $\bar{4}2mI$ & $\bar{4}m2I$ & $I\bar{4}m2$\\
        \cline{3-6}
        && $I\bar{4}c2$ & $\bar{4}m2I$ & $\bar{4}2mI$ & $Im\bar{4}2$\\
        \cline{2-6}
        &\multirow{2}{*}{$D_{4h}$}& $\begin{array}{c}P4/mcc,P4/nbm,P4/nnc,P4/mbm,\\P4/mnc,P4/nmm,P4/ncc,P4_2/mmc,\\P4_2/mcm,P4_2/nbc,P4_2/nnm,P4_2/mbc,\\P4_2/mnm,P4_2/nmc,P4_2/ncm \end{array}$ & $4/mmmP$ &$4/mmmP$&$P4/mmm$\\
        \cline{3-6}
        && $I4/mcm,I4_1/amd,I4_1/acd$ & $4/mmmI$ & $4/mmmI$&$I4/mmm$\\
        \hline
        \multirow{9}{*}{Trigonal}& $C_3$ &$P3_1,P3_2$ &$3P$ &$3P$&$P3$\\
        \cline{2-6}
        &\multirow{2}{*}{$D_3$}& $P3_121,P3_221$ & $321P$ &$312P$&$P312$\\
        \cline{3-6}
        && $P3_112,P3_212$ & $312P$ &$321P$&$P321$\\
        \cline{2-6}
        &\multirow{3}{*}{$C_{3v}$}&$P31c$ & $31mP$ & $3m1P$&$P3m1$\\
        \cline{3-6}
        && $P3c1$ & $3m1P$ & $31mP$&$P31m$\\
        \cline{3-6}
        && $R3c$ & $3mR$ &$3mR$&$R3m$\\
        \cline{2-6}
        &\multirow{3}{*}{$D_{3d}$} & $P\bar{3}1c$ & $\bar{3}1mP$ &$\bar{3}m1P$&$P\bar{3}m1$\\
        \cline{3-6}
        && $P\bar{3}c1$ & $\bar{3}m1P$&$\bar{3}1mP$&$P\bar{3}1m$\\
        \cline{3-6}
        && $R\bar{3}c$ & $\bar{3}mR$& $\bar{3}mR$ & $R\bar{3}m$\\
        \hline
        \multirow{7}{*}{Hexagonal}&$C_6$&$P6_1,P6_2,P6_3,P6_4,P6_5$& $6P$&$6P$&$P6$\\
		\cline{2-6}
		&$C_{6h}$ & $P6_3/m$ &$6/mP$ & $6/mP$&$P6/m$\\
        \cline{2-6}
        & $D_6$ & $P6_122,P6_222,P6_322,P6_422,P6_522$ & $622P$ &$622P$&$P622$\\
        \cline{2-6}
        &$C_{6v}$& $P6cc,P6_3cm,P6_3mc$&$6mmP$&$6mmP$ &$P6mm$\\
        \cline{2-6}
        &\multirow{2}{*}{$D_{3h}$}& $P\bar{6}2c$ & $\bar{6}2mP$ &$\bar{6}m2P$&$P\bar{6}m2$\\
        \cline{3-6}
        && $P\bar{6}c2$ & $\bar{6}m2P$ & $\bar{6}2mP$&$P\bar{6}2m$\\
        \cline{2-6}
        &$D_{6h}$ & $P6/mcc,P6_3/mcm,P6_3/mmc$ & $6/mmmP$ &$6/mmmP$&$P6/mmm$\\
        \hline
        & $T$& $I2_13$ & $23I$ &$23F$&$F23$\\
        \cline{2-6}
        &\multirow{3}{*}{$T_h$}& $Pa\bar{3},Pn\bar{3}$ & $m\bar{3}P$ &$m\bar{3}P$&$Pm\bar{3}$\\
        \cline{3-6}
        && $Ia\bar{3}$ &$m\bar{3}I$&$m\bar{3}F$&$Fm\bar{3}$\\
        \cline{3-6}
        && $Fd\bar{3}$ & $m\bar{3}F$ &$m\bar{3}I$&$Im\bar{3}$\\
        \cline{2-6}
        \multirow{6}{*}{Cubic}&\multirow{3}{*}{$O$}&$P4_132,P4_232,P4_332$ & $432P$ &$432P$&$P432$\\
        \cline{3-6}
        && $I4_132$ & $432I$ &$432F$&$F432$\\
        \cline{3-6}
        && $F4_132$ & $432F$&$432I$&$I432$ \\
        \cline{2-6}
        &\multirow{3}{*}{$T_d$} & $P\bar{4}3n$ & $\bar{4}3mP$ &$\bar{4}3mP$&$P\bar{4}3m$ \\
        \cline{3-6}
        && $I\bar{4}3d$ & $\bar{4}3mI$&$\bar{4}3mF$&$F\bar{4}3m$\\
        \cline{3-6}
        && $F\bar{4}3c$ & $\bar{4}3mF$&$\bar{4}3mI$&$I\bar{4}3m$\\
        \cline{2-6}
        &\multirow{3}{*}{$O_h$} & $Pn\bar{3}n,Pm\bar{3}n,Pn\bar{3}m$ & $m\bar{3}mP$&$m\bar{3}mP$&$Pm\bar{3}m$\\
        \cline{3-6}
        && $Ia\bar{3}d$ & $m\bar{3}mI$ &$m\bar{3}mF$&$Fm\bar{3}m$\\
        \cline{3-6}
        &&  $Fm\bar{3}c,Fd\bar{3}m,Fd\bar{3}c$ & $m\bar{3}mF$&$m\bar{3}mI$&$Im\bar{3}m$\\
        \hline
\end{longtable}}
\end{center}

\section{Realization of $k$-NSGs}
In this section, we begin by discussing the origin of projective representations from gauge fluxes over lattices. Subsequently, we demonstrate the realizations of specific $k$-NSGs through tight-binding models. These $k$-NSGs encompass all four 2D $k$-NSGs as well as several representative 3D $k$-NSGs.
\subsection{Projective representations from gauge fluxes over lattices}

Let us consider the Hilbert space of a single spinless particle hopping on lattice. The Hilbert space is spanned by $|i\rangle$ with $i$ labeling all lattice sites. The space group of the lattice is $SG$ consisting of lattice translations, rotations, reflections and their combinations. Each $g\in SG$ operates on the Hilbert space in the following way. First, $g$ permutes lattice sites according to the defining spatial operative meaning of $g$, which is described by a permutation matrix $\Gamma_{ij}^g$, i.e.,
\begin{equation}
    g:|i\rangle\mapsto |i\rangle\Gamma_{ij}^g.
\end{equation}

A quantum state $|\psi\rangle$ of the particle is described by a wave function $\psi_i=\langle i|\psi\rangle$ under the real-space basis $|i\rangle$. Then, the symmetry operator $S(g)$ of $g$ is given by 
\begin{equation}
    S(g)=\Gamma^g.
\end{equation}
It operates on wave functions by matrix multiplication, i.e., $S(g)\psi$. $S(g)$ is a representation of the space group $SG$ because for any $g_1,g_2\in SG$,
\begin{equation}
    S(g_1)S(g_2)=S(g_1g_2).
\end{equation}
Then, we further consider gauge transformations. After the spatial operation, a gauge transformation may be additionally applied, i.e., each $|i\rangle$ is multiplied by a phase factor $e^{i\theta_i^g}$. The gauge transformation is described by a diagonal matrix $\bm{G}_{ij}^g=e^{i\theta_i^g}\delta_{ij}$. Then, the symmetry operator for $g\in SG$ is modified as 
\begin{equation}
    \bm{S}(g)=\bm{G}^gS(g)=\bm{G}^g\Gamma^g.
\end{equation}
Then, the multiplication rules are modified as
\begin{equation}
    \bm{S}(g_1)\bm{S}(g_2)=\Delta(g_1,g_2)\bm{S}(g_1g_2),
\end{equation}
where 
\begin{equation}
    \Delta(g_1,g_2)=\bm{G}^{g_1}[\Gamma^{g_1}\bm{G}^{g_2}(\Gamma^{g_1})^{-1}](\bm{G}^{g_1g_2})^{-1}.
\end{equation}

It is significant to notice that $\Delta(g_1,g_2)$ is a diagonal matrix, because $\Gamma^{g_1}\bm{G}^{g_2}(\Gamma^{g_1})^{-1}$ is diagonal as a conjugate of a diagonal matrix by a permutation. 

We then further consider a one-particle tight-binding model
\begin{equation}
    \widehat{H}=\sum_{i,j}H_{ij}|i\rangle\langle j|
\end{equation}
that preserves the space group $SG$. Then, the symmetry constraints are given by
\begin{equation}
    [H,\bm{S}(g)]=0
\end{equation}
for all $g\in SG$. It is easy to see that 
\begin{equation}
    [\Delta(g_1,g_2),H]=0.
\end{equation}

Let us assume that any two any two sites can be connected by hopping process, i.e., the model corresponds to a connected map. Then,  $\Delta(g_1,g_2)$ is proportional to the identity matrix, i.e., $\Delta(g_1,g_2)=\nu(g_1,g_2)1$ with $\nu(g_1,g_2)\in U(1)$. Thus, we have the projective representation of the symmetry group in real space,
\begin{equation}
    \bm{S}(g_1)\bm{S}(g_2)=\nu(g_1,g_2)\bm{S}(g_1,g_2)
\end{equation}

Clearly, the (projective) representation in real space definitely not irreducible. To decompose the (projective) representation into irreducible representations, one can perform the Fourier transform, which is a unitary transformation that transforms the representation into momentum space. Then, in momentum space, (projective) representation can be further decomposed into irreducible components.

\subsection{Model realization of all 2D $k$-NSGs}
Now, we present the concrete model realizations of the four 2D $k$-NSGs one by one.

For the $k$-NSG $Pg$, the point group is $D_1=\{E,M_x\}$ with the action of $M_x$ on momentum space given by 
\begin{equation}
    M_x:\bm{k}\rightarrow(-k_x,k_y+\bm{G}_y/2),
\end{equation} 
where the fractional translation of $M_x$ is $\bm{\kappa}_{M_x}=\bm{G}_y/2$. The corresponding $r$-SSG is $Pm$, which can be expressed as  
\begin{equation}
    Pm=\langle e_x,e_y,M_x|[e_x,e_y],(M_xe_x)^2,[M_x,e_y],M_x^2\rangle.
\end{equation}

According to the Tab.\ref{tab:2D_k-NSGs}, to realize $Pg$ in momentum space, we need to take the multiplier of $Pm$ in real space as
\begin{equation}
    \nu[(\bm{t}_1,E),(\bm{t}_2,R)]=1,~~\nu[(\bm{t}_1,M_x),(\bm{t}_2,R)]=(-1)^{n_2^y},
\end{equation}
where $R=E$ or $M_x$ and $\bm{t}_i=n_i^xe_x+n_i^ye_y,i=1,2$. With this multiplier, the projective representation $\rho$ satisfies the following algebraic relation:
\begin{equation}\label{pgp}
       \rho(M_x)\rho(e_y)\rho(M_x^{-1})=-\rho(e_y),
\end{equation}
where the minus sign comes form the nontrivial multiplier $\nu$. 

\begin{figure}[!t]
    \centering
    \includegraphics[width=\linewidth]{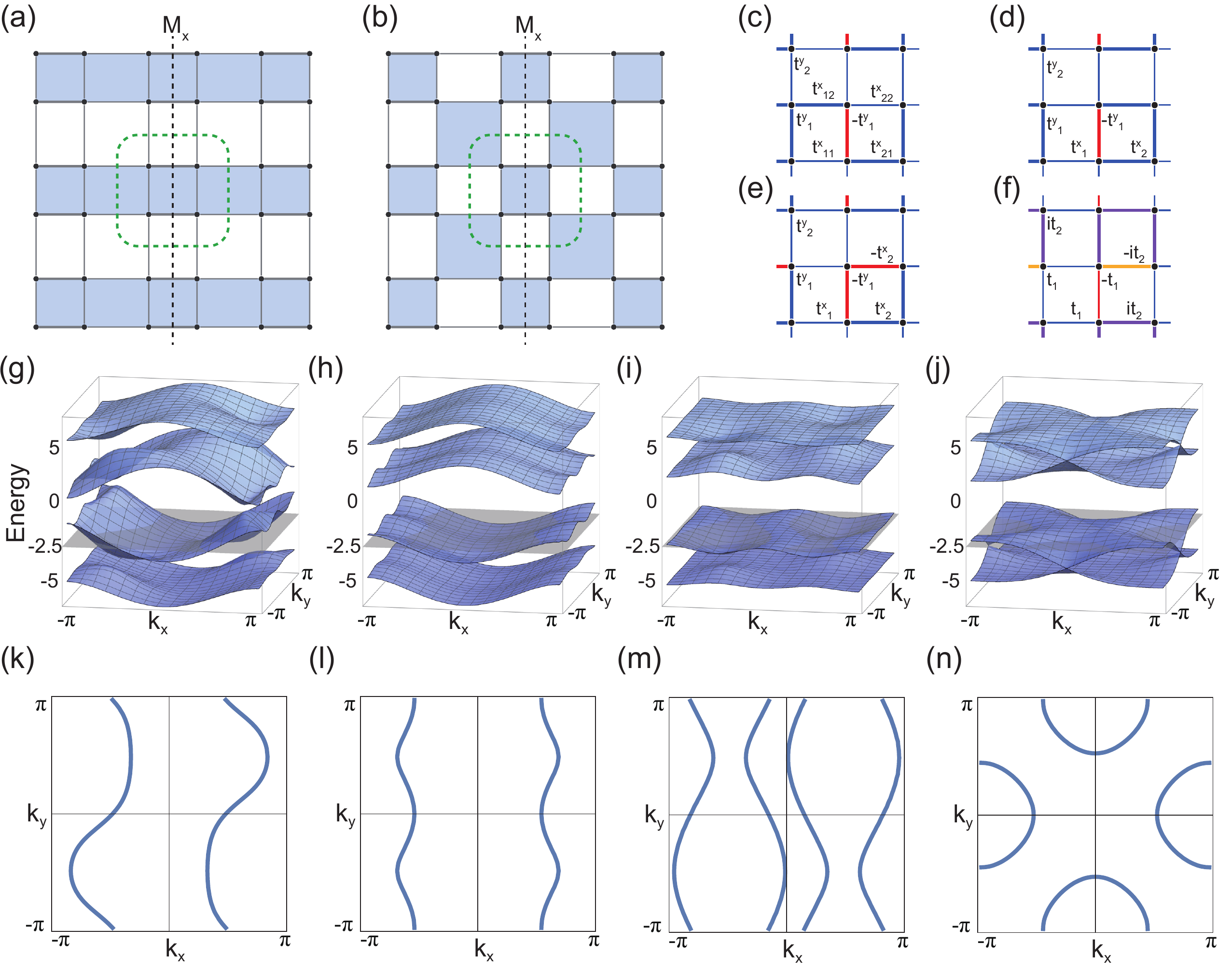}
    \caption{(a) The flux configuration which can realize $k$-NSG $Pg$ and $Pmg$. (b) The flux configuration which can realize $k$-NSG $Pg,Pgg$ and $P4g$. In (a) and (b), the blue shaded regions have $\pi$ flux through them and the chosen unit cells are specified by the green dashed regions. (c), (d), (e) and (f) labels the hopping amplitudes of lattice model of $k$-NSG $Pg, Pmg, Pgg$ and $P4g$ respectively. (g), (h), (i), (j) are energy spectrum of $k$-NSG $Pg, Pmg, Pgg, P4g$ model. (k), (l), (m) and (n) are constant energy cut of lattice model of $k$-NSG $Pg, Pmg, Pgg$ and $P4g$ at $E=-2.5$ respectively.}\label{sup_fig1}
\end{figure}

To realize this projective representation of $Pm$, we can start by designing a lattice with ordinary $Pm$ symmetry, and then introduce $\pi$ flux into every plaquette encircled by $M_xe_yM_x^{-1}e_y^{-1}$.  In Fig.~\ref{sup_fig1}(a), we provide one type flux configuration satisfying this requirement. Furthermore, in Fig.\ref{sup_fig1}(c), we demonstrate the hopping amplitudes that realize the projective representation of $Pm$ for this particular flux configuration.

The tight-binding model in Fig.~\ref{sup_fig1}(c) can be written as 
\begin{equation}
    \begin{split}
    \mathcal{H}_1&=\sum_{i}t_{11}^xa_{i,j}^{\dagger}d_{i,j}+t_{12}^xb_{i,j}^{\dagger}c_{i,j}+t_1^y(a_{i,j}^{\dagger}b_{i,j}-d_{i,j}^{\dagger}c_{i,j}) \\ 
    &+\sum_{i,j}t_{22}^xc_{i,j}^{\dagger}b_{i+1,j}+t_{21}^xd_{i,j}^{\dagger}a_{i+1,j}+t_2^y(b_{i,j}^{\dagger}a_{i,j+1}+c_{i,j}^{\dagger}d_{i,j+1})+h.c.,
    \end{split}
\end{equation}
A constant-energy contour of this model has been shown in Fig.~\ref{sup_fig1}(k), where the parameters are taken as $t^x_{11}=t^x_{22}=1,t^x_{12}=t^x_{21}=3.5,t^y_1=1,t^y_2=1.5,E=-2.5$. Clearly, the right line can be gotten by glide reflecting the left line  through $k_x=0$.

In the case of the $k$-NSG $Pmg$, the point group is $D_2=\{E,M_x,M_y,M_xM_y\}$, with the actions of $M_x,M_y$ on momentum space given by 
\begin{equation}
    \begin{split}
        &M_x:\bm{k}\rightarrow (-k_x,k_y+\bm{G}_y/2),\\
        &M_y:\bm{k}\rightarrow (k_x,-k_y),
    \end{split}
\end{equation}
where the fractional translations are $\bm{\kappa}_{M_x}=\bm{\kappa}_{M_xM_y}=\bm{G}_y/2$ and $\bm{\kappa}_{M_y}=0$. The corresponding $r$-SSG is $Pmm$, which can be presented as
\begin{equation}
    Pmm=\langle e_x,e_y,M_x,M_y|[e_x,e_y],(M_xe_x)^2,[M_x,e_y],(M_ye_y)^2,[M_y,e_x],M_x^2,M_y^2\rangle.
\end{equation}

To realize $k$-NSG $Pmg$, the multiplier of $Pmm$ in real space should be taken as
\begin{equation}
    \begin{split}
        &\nu[(\bm{t}_1,M_x),(\bm{t}_2,R)]=(-1)^{n_2^y},\\
        &\nu[(\bm{t}_1,M_y),(\bm{t}_2,R)]=0,
    \end{split}
\end{equation}
which also modifies the algebraic relation of the representation $\rho$ to 
\begin{equation}
        \rho(M_x)\rho(e_y)\rho(M_x^{-1})=-\rho(e_y).
\end{equation}
Thus, to realize the $k$-NSG $Pmg$, we can also adopt the flux configurations shown in Fig. \ref{sup_fig1}(a), only with hopping amplitudes satisfying projective $Pmm$ symmetry, as illustrated in Fig. \ref{sup_fig1}(d).  The model in Fig. \ref{sup_fig1}(d) can be obtained by setting $t^x_{11}=t^x_{12}=t^x_1,t^x_{21}=t^x_{22}=t^x_2$ in $\mathcal{H}_1$. Fig.~\ref{sup_fig1}(l) depicts a constant-energy contour of the model in Fig. \ref{sup_fig1}(d), where the parameters are taken as $E=-2.5,t_1^x=1,t^x_2=3.5,t^y_1=2,t^y_2=1.5$. These curves exhibit reflection symmetry through $k_y=0$ and glide reflection symmetry through $k_x=0$.

For the $k$-NSG $Pgg$, the point group is still $D_2$, but with the actions of $M_x,M_y$ on momentum space as
\begin{equation}
    \begin{split}
        &M_x:\bm{k}\rightarrow(-k_x,k_y+\bm{G}_y/2),\\
        &M_y:\bm{k}\rightarrow(k_x+\bm{G}_x/2,-k_y),
    \end{split}
\end{equation}
where fractional translations are $\bm{\kappa}_{M_x}=\bm{G}_y/2$ and $\bm{\kappa}_{M_y}=\bm{G}_x/2$. The corresponding $r$-SSG is still $Pmm$, but now the multiplier should be taken as 
\begin{equation}
    \begin{split}
        &\nu[(\bm{t}_1,M_x),(\bm{t}_2,R)]=(-1)^{n_2^y},\\
        &\nu[(\bm{t}_1,M_y),(\bm{t}_2,R)]=(-1)^{n_2^x},
    \end{split}
\end{equation}
which gives rise to
\begin{equation}
    \begin{split}
        &\rho(M_x)\rho(e_y)\rho(M_x^{-1})=-\rho(e_y),\\
        &\rho(M_y)\rho(e_x)\rho(M_y^{-1})=-\rho(e_x).\\
    \end{split}
\end{equation}

To realize the $k$-NSG $Pgg$, we need to adopt the flux configuration shown in Fig. \ref{sup_fig1}(a) so that $Pmm$ symmetry is satisfied and each plaquette circled by $M_xe_yM_x^{-1}e_y^{-1}$ or $M_ye_xM_y^{-1}e_x^{-1}$ has $\pi$ flux through it. For this type of flux configuration, we provide the hopping amplitudes that realize the projective representation of $Pmm$ in Fig. \ref{sup_fig1}(e).

The tight-binding model in Fig. \ref{sup_fig1}(e) can be written as 
\begin{equation}
    \begin{split}
        \mathcal{H}_2&=\sum_{i}t_{1}^x(a_{i,j}^{\dagger}d_{i,j}+b_{i,j}^{\dagger}c_{i,j})+t_1^y(a_{i,j}^{\dagger}b_{i,j}-d_{i,j}^{\dagger}c_{i,j}) \\ 
        &+\sum_{i,j}t_{2}^x(c_{i,j}^{\dagger}b_{i+1,j}-d_{i,j}^{\dagger}a_{i+1,j})+t_2^y(b_{i,j}^{\dagger}a_{i,j+1}+c_{i,j}^{\dagger}d_{i,j+1})+h.c..
        \end{split}
\end{equation}
Fig. \ref{sup_fig1}(m) shows a constant-energy contour of this model, where parameters are taken as $E=-2.5,t^x_1=t^y_1=1,t^x_2=3.5,t^y_2=1.5$. It is obvious that this figure has glide reflection symmetry along two directions.

For the $k$-NSG $P4g$, the point group is $D_4=\langle C,M_x|C^4,M_x^2,(CM_x)^2\rangle$, with the actions of $C$ and $M_x$ on momentum space given by 
\begin{equation}
    \begin{split}
        &C:\bm{k}\rightarrow (k_y,-k_x),\\
        &M_x:\bm{k}\rightarrow (-k_x+\bm{G}_x/2,k_y+\bm{G}_y/2),
    \end{split}
\end{equation}
where the fractional translation is $\bm{\kappa}_{M_x}=\bm{G_x}/2+\bm{G}_y/2$. The corresponding $r$-SSG is $P4m$, which can be presented as 
\begin{equation}
    P4m=\langle e_x,C,M|(e_xC)^4,(e_xC^2)^2,(e_xM_x)^2,(CM_x)^2,C^4,M_x^2 \rangle.
\end{equation}
The multiplier  should be taken as
\begin{equation}
    \begin{split}
        &\nu[(\bm{t}_1,C),(\bm{t}_2,R)]=0,\\
        &\nu[(\bm{t}_1,M_x),(\bm{t}_2,R)]=(-1)^{n_2^x+n_2^y}.
    \end{split}
\end{equation}
which gives rise to
\begin{equation}
    \begin{split}    
        &\rho(M_x)\rho(e_x)\rho(M_x^{-1})=-\rho(e_x^{-1}),\\
        &\rho(M_x)\rho(e_y)\rho(M_x^{-1})=-\rho(e_y).
    \end{split}
\end{equation}

To realize the $k$-NSG $P4g$, we can also adopt the flux configurations in Fig.~\ref{sup_fig1}(b), but with hopping amplitudes satisfying projective $P4m$ symmetry, as shown in Fig.~\ref{sup_fig1}(f). The four-fold rotation symmetry can be satisfied by taking $t^x_1=t^y_1=t_1,t^x_2=t^y_2=t_2$ in $\mathcal{H}_2$. Furthermore, we introduce imaginary hoppings to realize $[\rho(M_x)\rho(e_x)]^2=-1$. The Hamiltonian in momentum space can now be written as 
\begin{equation}
    \mathcal{H}_2(\bm{k})=\begin{bmatrix}
        0& t_1+it_2e^{-ik_y} &0&t_1-it_2e^{-ik_x}\\
        t_1-it_1e^{ik_y}&0&t_1+it_2e^{-ik_x}&0\\
        0&t_1-it_2e^{ik_x}&0&-t_1-it_2e^{-ik_y}\\
        t_1+it_2e^{ik_x}&0&-t_1+it_2e^{ik_y}&0           
    \end{bmatrix}
\end{equation}
Fig. \ref{sup_fig1}(n) shows a constant-energy contour of this model, where parameters are taken as $E=-2.5,t_1=1,t_2=2.5$. It exhibits an obvious $P4g$ symmetry.

\subsection{Model realization of representative 3D $k$-NSGs}

We now present two examples of 3D $k$-NSGs and their model realizations.

The first example is the $k$-NSG $P222_1$. The point group of $k$-NSG $P222_1$ is $D_2=\langle C_y,C_z|[C_y,C_z],C_y^2,C_z^2\rangle$, with the following actions:
\begin{equation}
    \begin{split}
        &C_y:\bm{k}\rightarrow (-k_x,k_y,-k_z+\bm{G}_z/2),\\
        &C_z:\bm{k}\rightarrow (-k_x,-k_y,k_z+\bm{G}_z/2).
    \end{split}
\end{equation}
The fractional translations are $\bm{k}_{C_y}=\bm{k}_{C_z}=\bm{G}_z/2$. The corresponding $r$-SSG is $P222$, which can be presented as
\begin{equation}
    \begin{split}
    P222=&\langle e_x,e_y,e_z,C_y,C_z|[e_x,e_y],[e_y,e_z],[e_x,e_z],(C_yL_x)^2,[C_y,L_y],\\
    &(C_yL_z)^2,(C_zL_x)^2,(C_zL_y)^2,[C_z,L_z],[C_y,C_z],C_z^2,C_y^2 \rangle.
    \end{split}
\end{equation}
The multiplier should be taken as
\begin{equation}
    \begin{split}
        &\nu[(\bm{t}_1,C_y),(\bm{t}_2,R)]=(-1)^{n_2^z},\\
        &\nu[(\bm{t}_1,C_z),(\bm{t}_2,R)]=(-1)^{n_2^z},
    \end{split}
\end{equation}
where $\bm{t}_i=n_i^xe_x+n_i^ye_y+n_i^ze_z$. This leads to the following relations:
\begin{equation}
    \begin{split}
        &\rho(C_y)\rho(e_z)\rho(C_y^{-1})=-\rho(e_z^{-1}),\\
        &\rho(C_z)\rho(e_z)\rho(C_z^{-1})=-\rho(e_z).\\
    \end{split}
\end{equation}

\begin{figure}[!t]
    \centering
    \includegraphics[width=\linewidth]{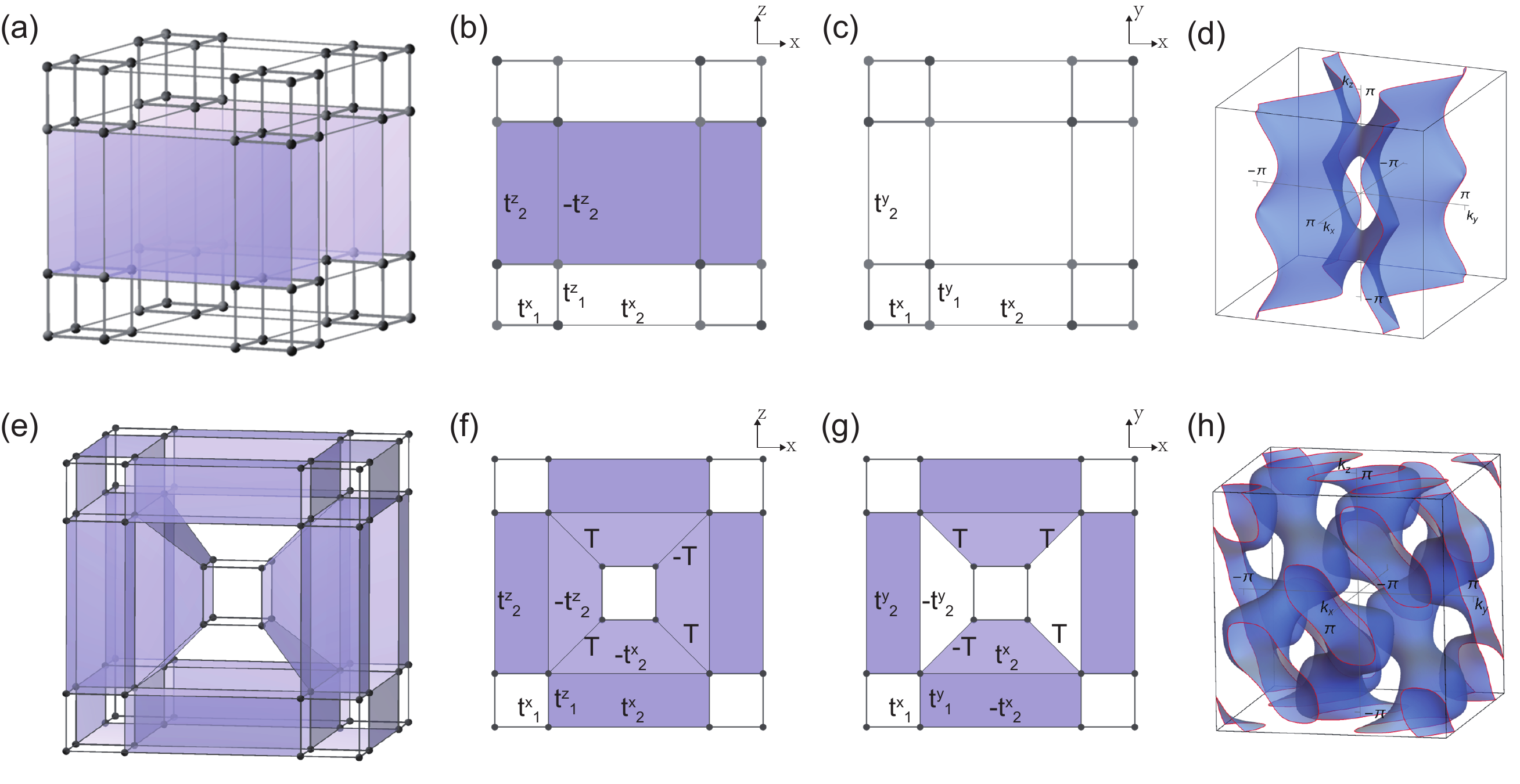}
    \caption{(a) and (e) are flux  realization of $k$-NSG $P222_1$ and $k$-NSG $Fddd$ respectively. In (a) and (e), all shaded regions has $\pi$ flux through them. (b) and (f) are front views of (a) and (b) respectively. (c) and (g) are top views of (a) and (b) respectively. (d) and (h) are constant-energy surfaces of models in (a) and (e) respectively.}\label{sup_fig2}
\end{figure}

The $k$-NSG $P222_1$ can be realized by the flux model shown in Fig.~\ref{sup_fig2}(a). In momentum space, the Hamiltonian of this tight-binding model can be written as 
$$
\setlength{\arraycolsep}{0.5pt}
\mathcal{H}_{P222_1}(\bm{k})=\begin{bmatrix}
        \epsilon&t^y_1+t^y_2e^{-ik_y}&0&t^x_1+t^x_2e^{-ik_x}&t^z_1+t^z_2e^{-ik_z}&0&0&0\\
        t^y_1+t^y_2e^{ik_y}&-\epsilon&t^x_1+t^x_2e^{-ik_x}&0&0&t^z_1+t^z_2e^{-ik_z}&0&0\\
        0&t^x_1+t^x_2e^{ik_x}&\epsilon&t^y_1+t^y_2e^{ik_y}&0&0&t^z_1-t^z_2e^{-ik_z}&0\\
        t^x_1+t^x_2e^{ik_x}&0&t^y_1+t^y_2e^{-ik_y}&-\epsilon&0&0&0&t^z_1-t^z_2e^{-ik_z}\\
        t^z_1+t^z_2e^{ik_z}&0&0&0&-\epsilon&t^y_1+t^y_2e^{-ik_y}&0&t^x_1+t^x_2e^{-ik_x}\\
        0&t^z_1+t^z_2e^{ik_z}&0&0&t^y_1+t^y_2e^{ik_y}&\epsilon&t^x_1+t^x_2e^{-ik_x}&0\\
        0&0&t^z_1-t^z_2e^{ik_z}&0&0&t^x_1+t^x_2e^{ik_x}&-\epsilon&t^y_1+t^y_2e^{ik_y}\\
        0&0&0&t^z_1-t^z_2e^{ik_z}&t^x_1+t^x_2e^{ik_x}&0&t^y_1+t^y_2e^{-ik_y}&\epsilon      
\end{bmatrix}
$$
where $t_i^{x(y,z)}$ are the hopping amplitudes labelled in  Fig.~\ref{sup_fig2}(b),(c) and $\epsilon$ is the on-site energy. The constant energy surface of this model, where parameters are taken as $E=-2,\epsilon=0.5,t^x_1=1,t^x_2=3.5,t^y_1=2,t^y_2=1.5,t^z_1=3,t^z_2=1$, is shown in Fig.~\ref{sup_fig2}(d).

The second example is $k$-NSG $Fddd$, which has been discussed in the main text. The basis of $\widehat{\mathcal{L}}$ is given by 
\begin{equation}
    \bm{G}_1=\pi(0,1,1),~\bm{G}_2=\pi(1,0,1),~\bm{G}_3=\pi(1,1,0).
\end{equation}
The point group of $k$-NSG $Fddd$ is $D_{2h}=\langle M_{k_1},M_{k_2},M_{k_3}|[M_{k_1},M_{k_2}],[M_{k_1},M_{k_3}],[M_{k_2},M_{k_3}] \rangle$, and the action of $M_{k_i}$ on momentum space is given by
\begin{equation}
    \begin{split}
        &M_{k_1}:\bm{k}\rightarrow (-k_x,k_y+\pi,k_z+\pi),\\
        &M_{k_2}:\bm{k}\rightarrow (k_x+\pi,-k_y,k_z+\pi),\\
        &M_{k_3}:\bm{k}\rightarrow (k_x+\pi,k_y\pi,-k_z).
    \end{split}
\end{equation}
The fractional translations are $\bm{\kappa}_{M_{k_i}}=\bm{G}_i/2,i=1,2,3$. The corresponding $r$-SSG is $Immm$, which can be presented as
\begin{equation}
    Immm=\langle e_{i},M_{k_j}|[e_{\alpha},e_{\beta}],M_{k_{\alpha}}e_{\alpha}M_{k_{\alpha}}^{-1}=e_1e_2e_3,M_{k_{\alpha}}e_{\beta}M_{k_{\alpha}}^{-1}=e_{\gamma}^{-1},M_{k_{\alpha}}^2\rangle,\quad i,j,\alpha,\beta,\gamma=1,2,3,\ \alpha\not=\beta\not=\gamma.
\end{equation}
The basis of $\mathcal{L}$ is given by 
\begin{equation}
    e_1=(-1,1,1),~e_2=(1,-1,1),~e_3=(1,1,-1)
\end{equation}
The multiplier components are given by 
\begin{equation}
    \nu[(\bm{t}_1,M_{k_i}),(\bm{t}_2,R)]=(-1)^{n_2^i},\quad i=1,2,3.
\end{equation}
where $\bm{t}_j=n_j^1e_1+n_j^2e_2+n_j^3e_3,~j=1,2$. This gives rise to the projective algebra 
\begin{equation}
    \begin{split}
        &\rho(M_{k_i})\rho(e_i)\rho(M_{k_i}^{-1})=-\rho(e_1+e_2+e_3),\\
        &\rho(M_{k_i})\rho(e_j)\rho(M_{k_i}^{-1})=\rho(e_k),\quad i\not=j\not=k.
    \end{split}
\end{equation}

The $k$-NSG $Fddd$ can be realized by the flux model shown in  Fig.~\ref{sup_fig2}(e). In momentum space, the Hamiltonian $\mathcal{H}_{Fddd}(\bm{k})$ of this model can be written as 
$$
    \begin{bmatrix}
        \begin{smallmatrix}
        0&t^y_1-t^y_2e^{-2ik_y}&0&t^x_1+t^x_2e^{-2ik_x}&t^z_1+t^z_2e^{-2ik_z}&0&Te^{-i(k_x+k_y+k_z)}&0\\
        t^y_1-t^y_2e^{2ik_y}&0&t^x_1-t^x_2e^{-2ik_x}&0&0&t^z_1-t^z_2e^{-2ik_z}&0&-Te^{-i(k_x-k_y+k_z)}\\
        0&t^x_1-t^x_2e^{2ik_x}&0&t^y_1+t^y_2e^{2ik_y}&-Te^{i(k_x+k_y-k_z)}&0&t^z_1+t^z_2e^{-2ik_z}&0\\
        t^x_1+t^x_2e^{2ik_x}&0&t^y_1+t^y_2e^{-2ik_y}&0&0&-Te^{i(k_x-k_y-k_z)}&0&t^z_1-t^z_2e^{-2ik_z}\\
        t^z_1+t^z_2e^{2ik_z}&0&-Te^{-i(k_x+k_y-k_z)}&0&0&t^y_1+t^y_2e^{-2ik_y}&0&t^x_1-t^x_2e^{-2ik_x}\\
        0&t^z_1-t^z_2e^{2ik_z}&0&-Te^{-i(k_x-k_y-k_z)}&t^y_1+t^y_2e^{2ik_y}&0&t^x_1+t^x_2e^{-2ik_x}&0\\
        Te^{i(k_x+k_y+k_z)}&0&t^z_1+t^z_2e^{2ik_z}&0&0&t^x_1+t^x_2e^{2ik_x}&0&t^y_1-t^y_2e^{2ik_y}\\
        0&-Te^{i(k_x-k_y+k_z)}&0&t^z_1-t^z_2e^{2ik_z}&t^x_1-t^x_2e^{2ik_x}&0&t^y_1-t^y_2e^{-2ik_y}&0         
    \end{smallmatrix}
    \end{bmatrix}
$$

Fig.~\ref{sup_fig2}(h) shows the constant energy surface of this model, where parameters are taken as $E=-8.3,t^x_1=1,t^x_2=3.5,t^y_1=2,t^y_2=1.5,t^z_1=1.5,t^z_2=2.5,T=0.5$. There is an obvious 1/4-translation invariance along diagonal direction at surface $k_z=\pi$.

\section{Experimental Realizations of gauge fluxes}
In this section, we present two possible methods that can be used to realize a $Z_2$ gauge field in engineering. Subsequently, we propose several experimental approaches to realize more general gauge fields.
\subsection{The dark-bright mechanism for engineering $Z_2$ Gauge Fields}
\begin{figure}[!h]
    \centering
    \includegraphics[width=\linewidth]{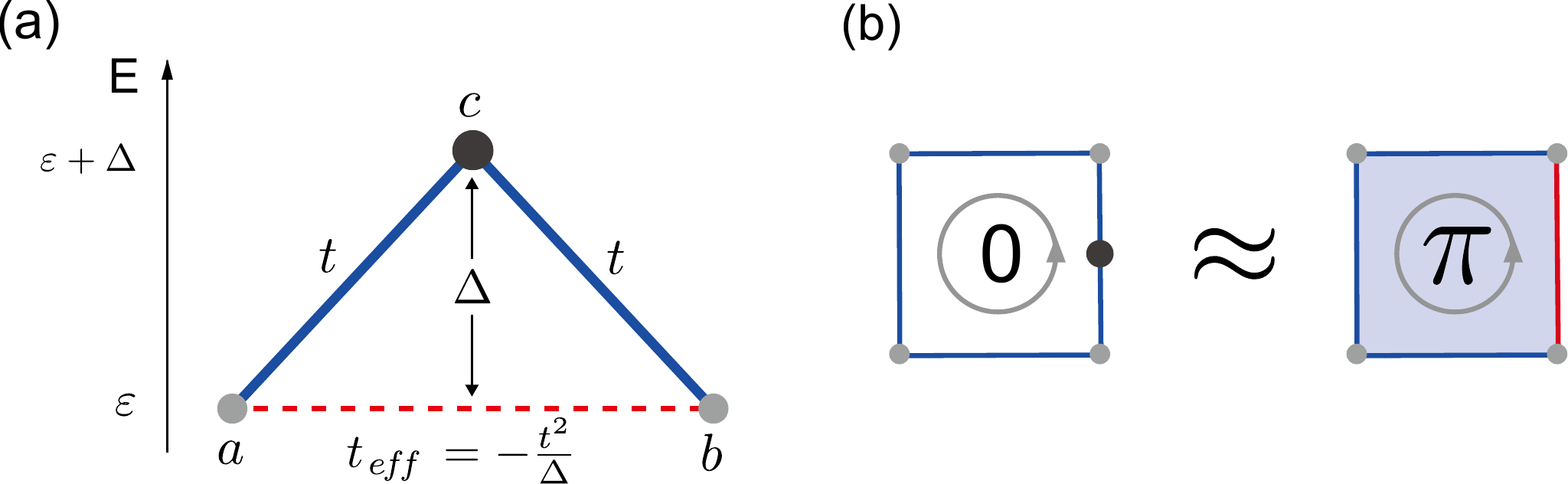}
    \caption{(a) The sites a,b has on-site energy $\epsilon$ and the site c has on-site energy $\epsilon+\Delta$. (b) The low-energy effective model of the left side is equivalent to the right side which has a $\pi$ flux. }\label{sub_fig3}
\end{figure}

Consider two sites a,b both with onsite energy $\epsilon$. By inserting an ancillary site c between them with onsite energy $\epsilon+\Delta$ shown in Fig.~\ref{sub_fig3}(a), the Hamiltonian is written as 
\begin{equation}
    H=\begin{bmatrix}
        \epsilon & 0 & t \\
        0 & \epsilon & t \\ 
        t & t & \epsilon+\Delta
    \end{bmatrix}
\end{equation}
in the basis $|a\rangle=(1,0,0)^T,|b\rangle=(0,1,0)^T,|c\rangle=(0,0,1)^T$. Here, the hoping amplitude $t$ between a,b and c is greater than 0. In the limit of $\Delta\gg\epsilon,t$, we have the eigenvalues and eigenvectors as 
\begin{equation}
    \begin{array}{ll}
        E_{<}^1=\epsilon, &|E_{<}^1\rangle= (1,-1,0)^T/\sqrt{2}, \\
        E_{<}^2=\epsilon-\frac{2t^2}{\Delta}, &|E_{<}^2\rangle\approx (1,1,\frac{2t}{\Delta})^T/\sqrt{2}, \\
        E_{>}=\epsilon+\Delta+\frac{2t^2}{\Delta},&|E_{>}\rangle \approx (\frac{t}{\Delta},\frac{t}{\Delta},1)^T.
    \end{array}
\end{equation}
Since $\Delta\gg\epsilon,t$, we can take $|E_{>}\rangle$ as high-energy excitation state. The low-energy subspace is spanned by $\{|E_{<}^{a}\rangle\}_{a=1,2}$. Then, in this subspace, we get the elements of the effective Hamiltonian
\begin{equation}
    [H_{eff}]_{ab}=\langle E_{<}^a|H|E_{<}^b \rangle,\quad a,b=1,2.
\end{equation}
By taking the approximation $|E_{<}^2\rangle\approx(1,1,0)^T/\sqrt{2}$ since $\frac{2t}{\Delta}\ll 1$, we have the low-energy effective Hamiltonian in the subspace of $\{|a\rangle,|b\rangle\}$ as
\begin{equation}
    H_{eff}=\begin{bmatrix}
        \epsilon-\frac{t^2}{\Delta}& -\frac{t^2}{\Delta}\\
        -\frac{t^2}{\Delta} & \epsilon-\frac{t^2}{\Delta}
    \end{bmatrix},
\end{equation}
in which emerges a $\pi$ flux with negative hopping amplitude $-\frac{t^2}{\Delta}$. As shown in Fig.~\ref{sub_fig3}(b), by inserting an ancillary site, the effective model of the plaquette with zero flux is equivalent to a plaquette with $\pi$ flux through it. And note that the fidelity is mainly dependent on $\Delta$.  

\subsection{Another mechanism for engineering $\mathbb{Z}_2$ gauge fields}
\begin{figure}[!h]
    \centering
    \includegraphics[width=\linewidth]{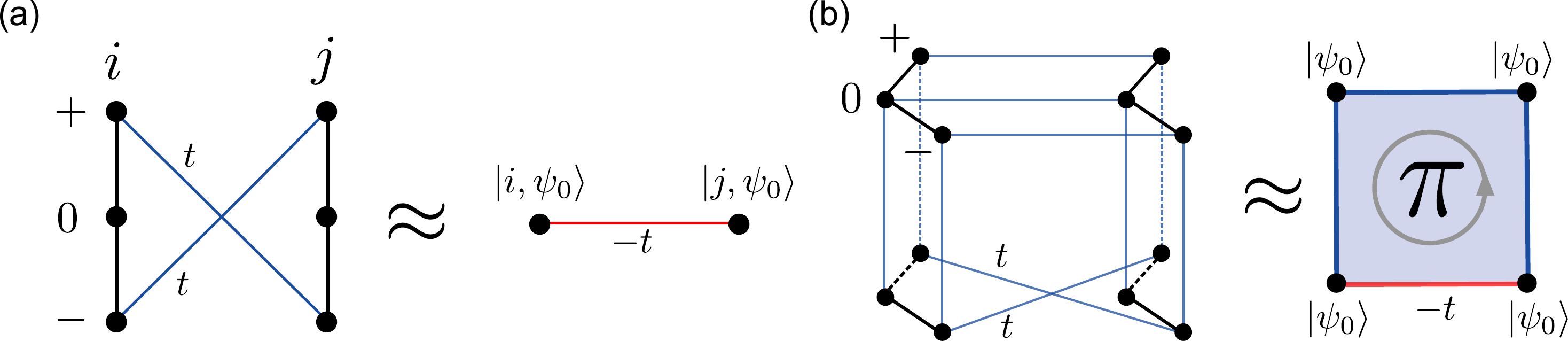}
    \caption{(a) Each cell has three sites which are denoted by $+,0,-$. There is an obvious chiral symmetry by exchanging $+$ and $-$. (b) Using structure in (a), we can construct a effective lattice with negative hopping amplitudes. On each cell, we only focus on zero energy model $|\psi_0 \rangle $.}\label{sup_fig4}
\end{figure}

Consider a unit cell consisting of three sites on the i-th lattice, the Hamiltonian of this subsystem can be written as
\begin{equation}
    \hat{h}_i=|i,+\rangle\langle i,0|+|i,-\rangle\langle i,0|+h.c..
\end{equation}
It has a zero-energy model due to the chiral symmetry. The eigenvector of the zero energy can be obtained as
\begin{equation}
    |i,\psi_0\rangle=(|i,+\rangle-|i,-\rangle)/\sqrt{2}.
\end{equation}
As shown in Fig.~\ref{sup_fig4}(a) , we consider a system with hopping between i-th cell and j-th cell. The transition operator can be written as
\begin{equation}
    \hat{T}=t(|i,+\rangle\langle j,-|+|j,+\rangle\langle i,-|+h.c.),
\end{equation}
where $t>0$ denotes the hopping amplitude. If we consider the effective hopping between zero energy models on i-th cell and j-th cell, 
\begin{equation}
    t_{ij}=\langle i,\psi_0|\hat{T}|j,\psi_0 \rangle=-t<0
\end{equation}
which can realize an negative hopping amplitude. So the effective model near zero energy can be used to construct a $\mathbb{Z}_2$ gauge field which is shown in the fig.~\ref{sup_fig4}(b).

In conclusion, we have presented two approaches for implementing the $\mathbb{Z}_2$ gauge field, both of which can be readily realized in engineering. Remarkably, out of the 157 known $k$-NSGs, 119 of them can be achieved using a $\mathbb{Z}_2$ gauge field. Consequently, these methods offer practical means to implement the majority of $k$-NSGs in engineering applications.

\subsection{Gauge fields in crystals}
Now, we give a brief review about gauge fields in crystalline systems besides the mechanisms discussed above. 

We first briefly review gauge fields in artificial systems, including cold atoms in optical lattices, photonic/acoustic crystals, periodic mechanical systems, electric circuit arrays below.
\begin{enumerate}[itemsep=2pt,topsep=0pt,parsep=0pt]
    \item In photonic crystals, the gauge field can be generated by modulation of the resonant frequencies, e.g., by adjusting the gap between site ring and link-ring wave guides \cite{ozawa2019topological, mittal2019photonic}.
    \item In acoustic crystals, $\mathbb{Z}_2$ hopping phases can be readily realized by coupling the resonators with wave guides on different sides \cite{Xue_2020, ma2019topological}.
    \item For cold atoms in optical lattices, we introduce two methods: rotating the optical lattice and laser-assisted tunneling \cite{zhang2018topological, cooper2019topological, dalibard2011colloquium}. i) Rotating optical lattice can introduce weak and uniform effective magnetic field and the side effect of Coriolis force should be compensated. ii) For the laser-assisted tunneling, the atomic hopping with desired gauge potentials can be engineered by coupling internal levels of atoms with laser beams. Different kinds of gauge fields can be induced, even the non-Abelian ones.
    \item For periodic mechanical systems, effective $\mathbb{Z}_2$ gauge field can be generated by tuning the stiffness coefficients of the spring connections \cite{prodan2009topological}.
    \item For electric circuit arrays, $\mathbb{Z}_2$ gauge fields can be realized by suitably choosing the capacitances and inductances \cite{imhof2018topolectrical, yu20204d, wang2023realization}.
\end{enumerate}
For condensed matter systems, we would like to emphasize an important fact, i.e., $ \Z_2 $ gauge fields preserve the time-reversal symmetry, which are essentially different from other $ U(1) $ gauge fields. Thus, $ \Z_2 $ gauge fields can be realized without introducing magnetism or magnetic fields. As such, $ \Z_2 $ gauge fields can be realized in a large class of condensed matter systems with preserved time reversal symmetry. We have discussed the so-called dark-bright mechanism above to achieve $ \Z_2 $ gauge fields. A well-known example is that in cuperates, the effective hopping amplitude between two $Cu$ sites (as mediated by the $O$ site in the middle) is negative.

For strongly correlated systems, there are emergent gauge fields in the low-energy effective theories. The $ \Z_2 $ gauge field, which defines the $ \Z_2 $ spin liquid, can naturally emerge in quantum spin liquid. In the mean-field theory of quantum spin liquid, close to the ground states the spinors are coupled to gauge field, particularly a $ \Z_2 $ gauge field as demonstrated in several works. Actually, perhaps it was the first time that physicists noticed the importance of the projective representations of space groups with a given gauge configuration, which leds to Xiao-Gang Wen's theory of PSG classification for quantum phases of spin liquids~\cite{Wen2002}. Another example is the Kitaev-type exactly solvable model~\cite{Kitaev2006}, where non-dynamical $ \Z_2 $ gauge fields are coupled with Majorana fermions.

\section{Other details}
\subsection{The 2D example model}
\begin{figure}[!h]
    \centering
    \includegraphics[width=\linewidth]{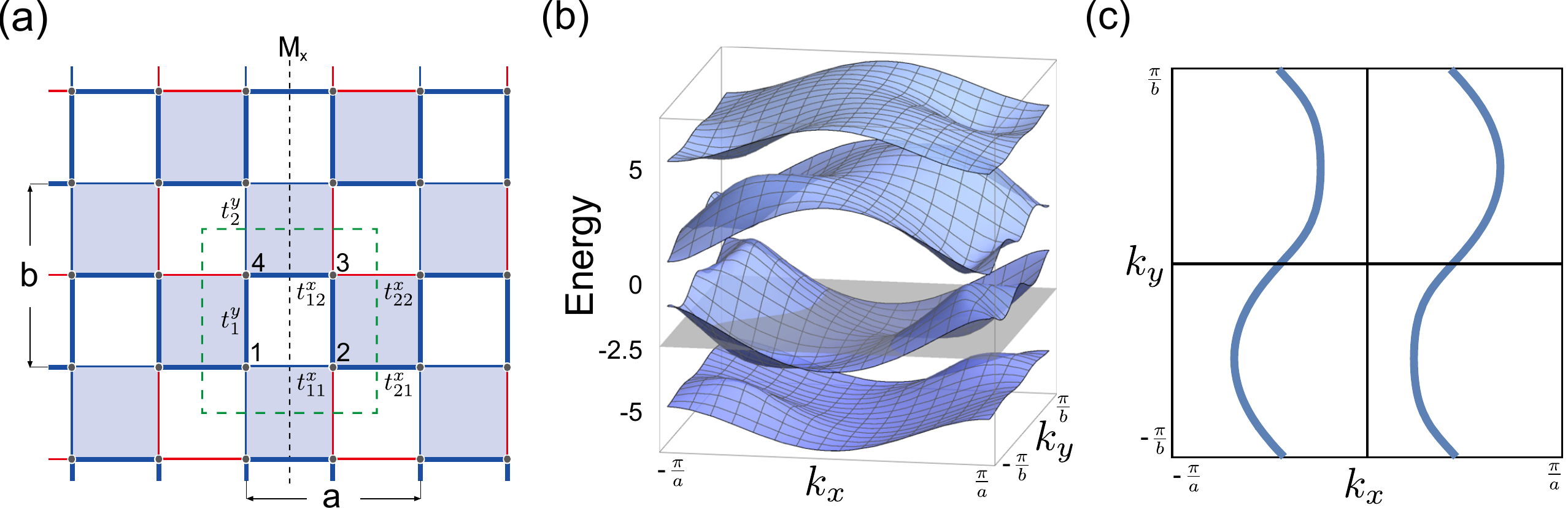}
    \caption{(a) The flux configuration and hopping amplitudes of a lattice model. The chosen unit cell is specified by the green dashed rectangle. The blue shaded regions have a $\pi$ flux(mod $2\pi$) through them. (b) The energy spectrum of $\mathcal{H}(\bm{k})$. (c) A constant energy cut of the energy spectrum at $E=-2.5$.}\label{sup_fig5}
\end{figure}

\begin{equation}
	\mathcal{H}(\bm{k})=\begin{bmatrix}
		\epsilon & t^x_{11}+t^x_{21}e^{-ik_xa} & 0 & t^y_1+t^y_2e^{-ik_yb}\\
		t^x_{11}+t^x_{21}e^{ik_xa} & \epsilon & t^y_1-t^y_2e^{-ik_yb} & 0\\
		0 & t^y_1-t^y_2e^{ik_yb} & -\epsilon & t^x_{21}-t^x_{22}e^{ik_xa}\\
		t^y_{1}+t^y_2e^{ik_yb} & 0 & t^x_{21}-t^x_{22}e^{-ik_xa} & -\epsilon 
	\end{bmatrix}
\end{equation}
Here, the hopping amplitudes $t^x_{ab},t^y_{c},a,b,c=1,2$ are labelled in Fig.~\ref{sup_fig5}(a), and $\pm\epsilon$ are the on-site energies. The unitary operator $U_{M_x}$ for the mirror operator is given by
\begin{equation}
	U_{M_x}=\begin{bmatrix}
		0&1&0&0\\
		1&0&0&0\\
		0&0&0&1\\
		0&0&1&0
	\end{bmatrix}.
\end{equation}
In the main text, we have presented the energy spectrum and constant energy cut in Fig.~\ref{fig:2D_model}(b), (c) by taking the parameters as $t^x_{11}=t^x_{22}=3.5,t^x_{12}=t^x_{21}=t^y_2=1,t^y_1=2.5,\epsilon=0.6$. Below, in Fig.~\ref{sup_fig5}, we also show the energy spectrum and constant energy cut with same parameters.

\subsection{Conjugation in projective representation}
The derivation detail is given as
\begin{equation}
	\begin{split}
		&\rho(\bm{t}',R)\rho(\bm{t},1)[\rho(\bm{t}',R)]^\dagger\\ =&\gamma(R\bm{t},R)\rho(\bm{t}'+R\bm{t},R)\rho(-R^T\bm{t}',R^T)/\gamma(-\bm{t}',R)\\
		=&\gamma(R\bm{t},R)\gamma(-\bm{t}',R)\rho(R\bm{t},1)/\gamma(-\bm{t}',R)\\
		=&\gamma(R\bm{t},R)\rho(R\bm{t},1)
	\end{split}
\end{equation}
Note that
\begin{equation}
	\rho(\bm{y},R)\rho(-R^T\bm{y},R^T)=\gamma(-\bm{y},R)\rho(0,1),
\end{equation}
and therefore $[\rho(\bm{y},R)]^\dagger=\rho(-R^T\bm{y},R^T)/\gamma(-\bm{y},R)$.

\subsection{Absence of $k$-NSGs in ordinary theory}

From the derivation above, it is now clear why $k$-NSGs do not exist in ordinary representation theory. For ordinary representations, $g(R\bm{t},R)=1$, and therefore $\rho(\bm{t},1)$ is sent to $\rho(R\bm{t},1)$. At $\bm{k}$ we have $\rho(\bm{t},1)=e^{i\bm{k}\cdot \bm{t}}$. Consequently, $e^{i\bm{k}\cdot \bm{t}}$ is sent to $e^{i\bm{k}\cdot R\bm{t}}=e^{i R^T\bm{k}\cdot \bm{t}}$, i.e., $\bm{k}\mapsto R^T \bm{k}$.

\subsection{Cocycle equation for real-space nonsymmorphic groups}
The equation can be seen from 
\begin{equation}
	\begin{split}
		(\bm{\tau}_{R_1}, R_1)(\bm{\tau}_{R_2},R_2) &=(\bm{\tau}_{R_1}+R_1\bm{\tau}_{R_2},R_1R_2)\\
		&=(\omega(R_1,R_2)+\bm{\tau}_{R_1R_2}, R_1R_2)
	\end{split}	
\end{equation}
where $\omega(R_1,R_2)\in \mathcal{L}$.

\end{document}